\documentclass[usenatbib]{mn2e}
\usepackage{graphicx}
\usepackage{amsbsy}
\usepackage{amssymb}

\def\HeI{He\,{\sevensize I}}
\def\HeII{He\,{\sevensize II}}

\makeatletter
\let\oldcr=\@tabularcr
\makeatother

\title[An asteroseismic signature of helium ionization] 
      {An asteroseismic signature of helium ionization} 
\author[G.\,Houdek \& D.\,O.\,Gough]
       {G.\,Houdek$^{1}$\thanks{e-mail: hg@ast.cam.ac.uk},
        D.\,O.\,Gough$^{1,2,3}$\\ 
       $^1$ Institute of Astronomy, Madingley Road, 
            Cambridge CB3 0HA, UK\\ 
       $^2$ Department of Applied Mathematics and Theoretical Physics, 
            Wilberforce Road, Cambridge CB3 0WA, UK\\
       $^3$ South African Astronomical Observatory, PO Box 9,
            Observatory 7935, South Africa}

\begin{document}

\maketitle

\begin{abstract} 
We investigate the influence of the ionization of helium on the
low-degree acoustic oscillation frequencies in model solar-type stars. 
The signature in the oscillation frequencies characterizing the 
ionization-induced depression of the first adiabatic exponent $\gamma$
is a superposition of two decaying periodic functions of frequency $\nu$, 
with `frequencies' that are approximately twice the acoustic
depths of the centres of the \HeI\ and \HeII\ ionization regions.  
That variation is probably best exhibited in the second frequency difference
$\Delta_2\nu_{n,l}\equiv\nu_{n-1,l}-2\nu_{n,l}+\nu_{n+1,l}$.
We show how an analytic approximation to the variation of $\gamma$
leads to a simple representation of this oscillatory contribution to
$\Delta_2\nu$ which can be used to characterize the $\gamma$ variation, 
our intention being to use it as a seismic diagnostic of the helium 
abundance of the star. We emphasize that the objective is to characterize
$\gamma$, not merely to find a formula for $\Delta_2\nu$ that reproduces
the data.
\end{abstract}


  \section{INTRODUCTION}
  \label{sec:intro}

The issue addressed in this paper concerns the direct interpretation of an
asteroseismic signature of helium ionization in terms of an
ionization-induced property of the stratification of the star that produces
that signature. The knowledge so obtained is important for designing
calibrations of theoretical stellar models against seismic data
(together with other astronomical data), for the purposes of obtaining
estimates of some of the gross properties of stars, such as their ages and
their chemical compositions. 

The first approach that might come to mind is to adjust apparently
appropriate parameters of the stellar models in such a way as to minimize 
some measure of deviation of the individual theoretical oscillation
eigenfrequencies from the observed frequencies. Without careful regard to the
extent to which those frequencies are influenced by the stellar
properties in mind, namely the helium abundance $Y$ in this
discussion, this naive procedure can hardly be optimal, although it has
been used: it was first carried out for the Sun, by \citet{cdg81}, 
who simply minimized the frequency misfit by least
squares. The procedure has the merit of being simple to execute, but
it is subject to all the uncertainties in stellar modelling, some of
which need not necessarily be very pertinent to the objective of the
calibration. It is more prudent to aim to work instead with certain
combinations of frequencies that are insensitive to the irrelevant
properties of the stellar modelling. Commonly used have been the
so-called large and small frequency separations, which are relatively
insensitive to the uncertain state of the outer layers of the stars
\citep[e.g.][]{cdg84, g86b}. \citet[][see also \citealp{u86, g87}]{cd86} 
illustrated how from a knowledge of these separations one
could in principle estimate the mass and age of a main-sequence star, 
provided that
its chemical composition were known; indeed, in the early days of
helioseismology these separations were often preferred to the raw
frequencies for solar calibrations \citep{cdg80, sng83, ur83}. But the
chemical composition of a star, particularly the helium abundance, is
often not well known, and it is therefore incumbent upon us to devise
more reliable ways of determining it, especially in view of the
high-precision data anticipated from the imminent space missions COROT
(Convection Rotation and Planetary Transits; \citealt{bagl03}) and
Kepler \citep{betal03, bbk05}, and from the ground-based campaigns
such as SONG (Grundahl et al. 2006) and those of the kind organized by
Kjeldsen, Bedding and their colleagues (e.g. Kjeldsen et
al. 2005). Even in the Sun, in which helium was discovered by
J.N. Lockyer and P.J.C. Janssen in prominences observed during the
eclipse of 1868, the spectroscopic determination of $Y$ remains
elusive to this day. This comes about because helium spectrum lines
are formed well above the photosphere, in the much hotter chromosphere
and corona where the deviation from local thermodynamic equilibrium
can be severe, and where the helium abundance is not even necessarily
the same as in the layers beneath the visible surface.  

In principle, a direct seismological signature of $Y$ in the envelope of a
(sufficiently cool) star can be sought through its effect on the sound
speed, via the depression of the value of the first adiabatic exponent
$\gamma$ induced by ionization, particularly if the depression occurs in
an essentially adiabatically stratified region of a convection zone.
The variation is seismically abrupt,
in the sense that, at least for lower-frequency modes, its radial
extent is less than or comparable with the inverse radial wavenumber
of the eigenfunctions. We call such a feature an acoustic glitch. It
causes a small shift in the eigenfrequencies, relative to those in a
corresponding putative star with no ionization, that shift being an
oscillatory function of the frequency itself. The amplitude of that
function increases monotonically with $Y$, and should therefore be a
robust indicator of the value of $Y$; the frequency of the function is
determined by the acoustic depth of the centre of ionization beneath
the seismic surface of the star.

Oscillatory components are produced also by other acoustic glitches,
such as that at the base of the convection zone which is caused by a
near-discontinuity in the gradient of the density scale height. They
are evident in the large separations
$\Delta_1\nu_{n,l}\equiv\nu_{n,l}-\nu_{n-1,l}$ between the cyclic
eigenfrequencies $\nu_{n,l}$ of seismic oscillations of order $n$ and
degree $l$, complicating the measurement of the underlying smooth
component of $\Delta_1\nu_{n,l}$, and also of other diagnostics such
as the small separation $\nu_{n,l}-\nu_{n-1,l+2}$.  Not taking them
into account leaves an undulation in the outcome of data-fitting when
plotted against the extremities of the range of frequencies employed
\citep{g01}, which could lead to misinterpretation. Therefore it is
prudent to separate the smooth and the oscillatory components, using
the two of them as complementary diagnostics. We address in this paper
a way in which that separation could be carried out, by analysing
signatures suggested by asymptotic analysis of the eigenfrequencies of
a grid of solar models, with the intention of determining parameters
that are good measures of the variation of the stratification in the
second helium ionization zone that are not unduly contaminated by
other properties.

Direct seismological signatures of the helium abundance in the solar
interior have been constructed from intermediate-degree acoustic modes by
various techniques \citep[e.g.][]{g84a, dgkt91, kcddgt92, vbp92,
phcd94b, ba95}.  In distant stars, however, only low-degree modes can
be observed, and we need to find a diagnostic in only those modes, a
need which has been stressed before by e.g. \citet{g90, g98}. 
One procedure is to attempt a direct inversion of low-degree modes for
the helium abundance $Y$, using the constraint that the bulk of the
convection zone is adiabatically stratified \citep{kos93}.
More commonly, the outer phase function $\tilde\alpha(\omega)$ in the
asymptotic relation 
\begin{equation}
(n+\tilde\alpha)\pi/\omega=\tilde F(w)+...
\label{eq:asympfrequ}
\end{equation}
\citep{g84b, g93, vor88, vbp91} has been used, for it
exhibits explicitly the oscillatory components of $\omega$ induced by 
ionization \citep[e.g.][]{bm90a, bm90b, phcd92, rv94}; here $\omega$ 
is the angular frequency of a mode of order $n$ and degree $l$,
$w=\omega/(l+\frac{1}{2})$ and $\tilde F$ is a function of $w$ which
depends also on the structure of the star. The idea is to determine
$\tilde\alpha$ by fitting the functional form of its asymptotic
relation to the seismic data (cf. \citealt{g86a, vor88, gv95}). It is
evident from that relation that $\tilde\alpha/\omega$ (and $\pi^{-1}\tilde F$) 
can be determined from data $\omega_{n,l}$ only up to an unknown additive
constant, so \citet{bv87} proposed using the derivative
$\tilde\beta={\rm d}(\omega^{-1}\tilde\alpha)/{\rm d}\omega$, which 
contains the
same information and is more elegant. The oscillatory component can be
largely separated by appropriate filtering (cf. \citealt{phcd94a}), and
the contribution from helium ionization used to calibrate stellar
models to determine $Y$ \citep[e.g.][]{phcd94b, phcd98, ltmg97, lg01}.

An alternative procedure is to consider the second frequency
difference, $\Delta_2\nu_{n,l}$ \citep{g90}, or even higher-order
differences \citep{ban94, b97}, which are much simpler to extract from
the data than the phase functions $\tilde\alpha$ and $\tilde\beta$ yet
appear to contain the same information.  It has been found expedient
to model their oscillatory components with parametrized functions
derived from representations of the acoustic glitches, and then to
determine the sought-for properties of the star by calibrating the
parameters corresponding to a
sequence of stellar models against the seismic data. Various
approximate formulae for the seismic signatures that are associated
with the helium ionization have been suggested and used, by
\citet{mt98, mt05}, \citet{g02}, \citet{mig03}, \citet{b04},
\citet{ver04} and \citet{btc04}, not all of which are
derived directly from explicit acoustic glitches. Gough used an
analytic function for modelling the dip in the first adiabatic
exponent. In contrast, \citet{mt98} assumed a triangular form, which
was adopted also by \citet{mt05}, \citeauthor{mig03} and Verner\,et\,al. 
These are the only two attempts to relate the low-degree seismic
signature directly to the properties of the ionization zone, although
the latter appears to be too crude to capture the physics 
adequately \citep{mt05}. 
\citet{b04}, Ballot\,et\,al.\,(2004) and \citet{pbt05} have adopted a 
seismic signature for helium ionization that is similar to that arising 
from a single discontinuity
in the sound speed \citep[see also][]{ban94,b97}, which is even less
appropriate; the artificial discontinuities in the sound speed and its
derivatives that this and the triangular representations possess cause 
the amplitude of
the oscillatory signal to decay with frequency too gradually, although
that deficiency may not be immediately noticeable within the limited 
frequency range in which adequate asteroseismic data are or will 
imminently be available. However, any inappropriate representation 
of the seismic signature can hardly be trusted to measure the 
characteristic of the glitch that causes it. 

Although the seismic quantities $\tilde\alpha, \tilde\beta$ associated with
low-degree modes and 
$\Delta_k\nu_{n,l}$ have been analysed in such a way as to enable the
magnitude of the oscillatory components produced by helium ionization to be
obtained, in none of the studies, save those of \citet{mt98, mt05} and
\citet{g02}, was the signature related directly to the form of the acoustic
glitch that produced them. They were simply used as a medium for fitting
theoretical models to seismic data, in the hope that the stellar properties
of interest, such as $Y$, had been determined reliably. It has not been 
demonstrated whether the inferences are essentially uncontaminated by other
properties of the star. Our aim in this paper is to design a seismic signature
that truly reflects the properties of the seismic glitches we wish to measure.
And we report below that we have been reasonably successful. The next step in
advancing our understanding would require a further independent study of 
stellar structure (and the equation of state) to ascertain how the glitch 
properties relate to $Y$.
The glitch amplitude increases with increasing $Y$. However, 
we have found that the rate at which it does so depends also on other factors,
such as the entropy of the adiabatically stratified ionization zones, as has
been noted previously by \citet{dgt88}, \citet{ltmg97} and \citet{phcd98}, and 
that therefore the glitch amplitude alone is not a simple measure of~$Y$.

In Sun-like stars, the stratification in the helium ionization zones
is very close to being adiabatic (and Reynolds stresses are
negligible), and consequently changes $\delta \gamma$ in 
$\gamma$ produce well defined
changes in the variation of sound speed with radius. We note, in
passing, that the ionization of hydrogen produces much greater
$\gamma$ variation, but unfortunately (for this study) the hydrogen
ionization zone is not even approximately adiabatically stratified
throughout; the form of the seismic glitch must be susceptible to the
uncertain formalism adopted to model the convective energy and
momentum fluxes, and relating it to $\gamma$ must therefore be
unreliable. For that reason we concentrate on the ionization of
helium, as have others before us.  We represent the variation of
$\gamma$ with a pair of Gaussian functions. This correctly results in
a decay of the amplitude of the seismic signature with oscillation
frequency that is faster than that which the triangular and the
single-discontinuity approximations imply, and also takes some account of
the two ionization states of helium.

The plan of the paper is as follows:
in the following Section we introduce and derive a simple seismic
diagnostic of a Gaussian acoustic glitch produced by only the (dominant) 
second ionization of helium,
together with a representation of the contribution from the base of the
convection zone, the latter being derived in Appendix\,\ref{sec:delomc}; 
the variation of the eigenfunction phase $\psi$ with acoustic depth $\tau$ 
is first represented by the simple formula $\psi\simeq\omega\tau+\epsilon$, 
where $\epsilon$ is taken to be constant. This diagnostic is
tested against artificial frequency data computed from two sequences of 
theoretical solar models, which we describe in Section\,\ref{sec:testing}.
Although the diagnostic can be fitted to the theoretical eigenfrequencies
tolerably well, the inferred value of the acoustic depth $\tau_{\rm II}$
of the glitch is too low, and the inferred amplitude 
$-\delta\gamma/\gamma\vert_{\tau=\tau_{\rm II}}$ is too high. This is as one 
might expect, partly because in this initial phase of the investigation we
have adopted only a single Gaussian function to represent the two ionization
stages of helium, and partly because we have overestimated 
$\psi(\tau_{\rm II})$ by not taking account of the acoustic cutoff frequency
$\omega_{\rm a}$. We rectify this, in two steps, in Section~4, first by
incorporating $\omega_{\rm a}$ into $\psi$, which essentially halves the
discrepancy between the true $\tau_{\rm II}$ and its seismologically 
inferred value, and also reduces the amplitude discrepancy considerably
(although it does not reduce the measure $\chi^2$ of the data misfit),
and then by adding in a simple way a contribution to the diagnostic
from the first stage of ionization of helium, which reduces the 
$\Delta_{\rm II}$ discrepancy by a factor three (and also reduces
$\chi^2$).
We must point out, however, that the 
value we must adopt for the ratio $\Delta_{\rm I}/\Delta_{\rm II}$ of the
characteristic widths of the \HeI\ and \HeII\ glitches in order to obtain
a good fit to the artificial data is rather larger than that suggested by
the models. We do not know why, although a change of this kind is perhaps
suggested
by the functional form of the contribution to the depression of $\gamma$ due
to the ionization of hydrogen, which is depicted in Fig.\,7.
We discuss our analyses further
in Section\,\ref{sec:discussion}, and we draw our
conclusions in Section\,\ref{sec:summary}.

   \section{The seismic diagnostic}
   \label{sec:seismodel}

A convenient and easily evaluated measure of the oscillatory component
is the second difference with respect to order $n$ of the
cyclic frequencies $\nu_{n,l}$ of like degree $l$:
\begin{equation}
\Delta_2\nu_{n,l}\equiv\nu_{n-1,l}-2\nu_{n,l}+\nu_{n+1,l}\,,
\label{eq:secdiff}
\end{equation}
\citep{g90}. This measure is contaminated less than the first
difference $\Delta_1\nu_{n,l}\equiv \nu_{n,l}-\nu_{n-1,l}$, otherwise
known as the large frequency separation, by the
smoothly varying components of $\Delta_1\nu$ (here and henceforth we
simplify the notation, where it is not ambiguous to do so, by omitting the
subscripts $n,l$). And it is less susceptible to data errors than the
fourth and other higher-order differences 
(see Appendix\,\ref{sec:DOGcallsthisAppxC}); moreover, adopting higher-order
differences risks requiring more consecutive modes than might be available
to provide a practical diagnostic.

Any localized region of rapid variation of the
propagation speed of an acoustic wave, which here we call an acoustic
glitch, induces an oscillatory component in $\Delta_2\nu$ with a
`cyclic frequency' approximately equal to twice the acoustic depth
\begin{equation}
\tau=\int_{r_{\rm glitch}}^R c^{-1}\,{\rm d}r
\end{equation}
of the glitch, where $r$ is a radial coordinate, $c$ is the adiabatic
sound speed and $R$ is the radius of the seismic surface of 
the star (essentially the surface on which $c^2$, if extrapolated 
linearly outwards from the outer layers of the adiabatically stratified
region of the convection zone, would vanish -- cf.~\citealt{bg90}).
The amplitude of the oscillatory component depends on the amplitude 
of the glitch, and
decays with $\nu$ once the inverse radial wavenumber of the mode
becomes comparable with or greater than the radial extent of the
glitch.  By calibrating a theoretical representation of the effect of
glitches against the observations one can, in principle, learn about
the characteristics of those glitches. Because only low-degree modes
are used, the $l$-dependence can safely be ignored, even down to the
base of the convection zone (except, possibly, in very-low-mass, almost
fully convective stars). We remark further on that in 
Section\,\ref{sec:discussion}.\\ 
If a frame of reference exists in which the
basic structure of a star, which in general is presumed to be rotating 
(but only slowly), is independent of time, then one can seek linearized
adiabatic oscillations in that frame whose time dependence is
sinusoidal with angular frequency $\omega$. The adiabatic oscillation
frequency $\omega$ is related to the displacement eigenfunction $\bxi$
through a variational principle~\citep{c63,lo67},
the derivation of the acoustic signature from which we now address.
We consider the structure of the star to
be described by smoothly varying functions upon which are superposed
glitches, each varying on a small spatial scale.  The contribution
$\delta\omega$ from those glitches to the eigenfrequency $\omega$ of a
mode of oscillation (with vanishing pressure gradient at the surface),
can then be written (see Appendix\,\ref{sec:varprin})
\begin{equation}
\delta\omega\simeq\frac{\delta{\cal K}-\omega\delta {\cal I}}{2\omega
{\cal I}}\,,
\label{eq:varprin3}
\end{equation}
where $\cal I$ is the mode inertia and $\delta{\cal K}$ depends on
the displacement eigenfunction $\bxi$ and on the glitches in the
equilibrium quantities, such as the (localized) perturbation
$\delta\gamma$ to the first adiabatic exponent $\gamma=(\partial\ln
p/\partial\ln\rho)_s$, $s$ being specific entropy, caused by
ionization.  In equation\,(\ref{eq:varprin3}), contributions from the
surface (which at most contribute a relatively smoothly varying
component to $\Delta_2\nu$) have been ignored.

We separate $\delta\omega$ into a smoothly varying component
$\delta_{\rm sm}\omega$ and an oscillatory component $\delta_{\rm
osc}\omega$:
\begin{equation}
\delta\omega=\delta_{\rm sm}\omega+\delta_{\rm osc}\omega\,,
\label{eq:delomega}
\end{equation}
(and we adopt a similar notation for other variables) and we approximate,
with a suitable eigenfunction normalization, the oscillatory component
$\delta_{\gamma,{\rm osc}}\omega$ associated with helium ionization as
follows:
\begin{displaymath}
\delta_{\gamma,{\rm osc}}\omega={\rm osc}(\delta_\gamma\omega)~,
\end{displaymath}
where
\begin{equation}
\delta_\gamma\omega:=
                   {\delta_\gamma{\cal K}\over2\omega \cal I}\,,
\label{eq:delomegaosc}
\end{equation}
with
\begin{equation}
\delta_\gamma{\cal K}\simeq\delta_\gamma{\cal K}_1 =
     \int(\delta\gamma)\,p({\rm div}\bxi)^2r^2\,{\rm d}r\,,
\label{eq:K1}
\end{equation}
(cf. equations\,(\ref{eq:A5}),(\ref{eq:A6})) and
\begin{equation}
{\cal I}\!=\!\int\!\rho\,{\bxi}\!\cdot\!{\bxi}\,r^2\,{\rm d}r\,,
\label{eq:inertia}
\end{equation}
the integrations being over the entire extent of the star, where $p$
and $\rho$ are pressure and density of the equilibrium state, and
$\delta\gamma$ is a suitably defined rapidly varying perturbation to
$\gamma$ induced by ionization; osc denotes oscillatory part. We have
confirmed numerically that the terms neglected in $\delta{\cal K}$ are
indeed substantially smaller than $\delta_\gamma{\cal K}$ (although they
may not be wholely negligible, as is evinced by the differences in
$\delta\gamma /\gamma\vert_{\tau_{\rm II}}$ in 
Figs~\ref{fig:gamma2_nine-fit} and \ref{fig:ff3}). 
In view of the variational principle, we have approximated the
eigenfrequencies by those of a corresponding nonrotating smoothly
varying (i.e. having no glitches) model star; consequently in
equations (\ref{eq:K1}) and (\ref{eq:inertia}) we have been able
to adopt the
usual separation of the components of the eigenfunctions $\bxi$ in
terms of functions of $r$ and spherical harmonics, carry 
out the angular integrations, and regard $\bxi$ to be a function
of $r$ alone. In the asymptotic limit of high-order
modes the (suitably normalized) radial component $\xi$ of the displacement
eigenfunction $\bxi$ and the associated Lagrangian pressure 
perturbation $\tilde{p}$ are given by\,\citep[e.g.][]{g93}
\begin{equation}
\xi\simeq\frac{K^{1/2}}{r\rho^{1/2}}\cos\psi\,,\qquad\qquad
\tilde{p}\simeq\frac{\omega^2\rho^{1/2}}{rK^{1/2}}\sin\psi\,,
\label{eq:8}
\end{equation}
where the phase function $\psi$ is given by
\begin{equation}
\psi=\int_r^{r_{\rm t}}K(r^\prime)\,{\rm
d}r^\prime+\frac{\pi}{4}\simeq\omega\tau+\epsilon\,,
\label{eq:phasefunc}
\end{equation}
in which here we have approximated the radial component of the wavenumber
of high-order modes as $K\simeq\omega/c$,
and we have assumed that $r$ is not close to $r_{\rm t}$,
the upper turning point at which the mode is reflected; $\epsilon$ is a
phase constant (strictly speaking, a phase variable, for it varies slowly 
with $\omega$, but in the simple approach that we adopt in just this 
section that variation is ignored) 
which takes some account of the fact that $K$ deviates
substantially from $\omega/c$ near the upper turning point. 
It follows that 
\begin{equation}
({\rm div}\bxi)^2=\left(\frac{\tilde{p}}{\gamma p}\right)^2
            \simeq\frac{\omega^3}{\gamma pcr^2}\sin^2\psi\,,
\end{equation}
and therefore equation\,(\ref{eq:K1}) can be written in terms of
\begin{equation}
\delta_{\gamma,{\rm osc}}{\cal K}_1\simeq
    -{1\over 2}\omega^3
     \int{\delta\gamma\over\gamma}\cos2(\omega\tau+\epsilon)\,{\rm d}\tau\,,
\label{eq:11}
\end{equation}
in which the limits of integration are such as to include the region
in which $\delta\gamma$ is nonzero; moreover
\begin{equation}
{\cal I}\simeq\omega\int^T_0\cos^2(\omega\tau+\epsilon)\,{\rm d}\tau
 \simeq{1\over2}T\omega\,,
\label{eq:inertia1}
\end{equation}
where $T$ is the acoustic radius of the star (i.e. the sound travel
time from the acoustic surface to the centre of the star). Note that
the normalization of this expression for the inertia differs from the
usual. The neglect of the acoustic cutoff frequency in the
expression for $K$ will be dropped in Section 4.1. However, we shall
continue to neglect the degree-dependence, which is a good
approximation for the low-degree modes detectable by asteroseismology;
we justify that neglect in Section 5.  

In this introductory discussion we represent the acoustic glitch
induced by the (second) ionization of helium by a Gaussian function
about the acoustic depth $\tau=\tau_{\rm II}$ of the centre of the
\HeII\ ionization region (see Fig.~\ref{fig:gamma_fit}), as did \citet{g02}:
\begin{equation}
{\delta\gamma\over\gamma}\simeq
G(\tau;\Gamma_{\rm II},\tau_{\rm II},\Delta_{\rm II})\equiv
-{1\over\sqrt{2\pi}}{\Gamma_{\rm II}\over{\Delta_{\rm II}}}
                {\rm e}^{-(\tau-{\tau_{\rm II}})^2/2{\Delta_{\rm II}}^2}\,,
\label{eq:dgog}
\end{equation}
in which $\Gamma_{\rm II}$ and $\Delta_{\rm II}$ are constants and
$\Delta_{\rm II}$ is much smaller than both $\tau_{\rm II}$ and
$T-\tau_{\rm II}$. We improve on this simple expression in Section
4.2. The oscillatory component, $\delta_{\gamma,{\rm osc}}\omega$,
imparted by the glitch can then be estimated asymptotically from
equations (\ref{eq:delomegaosc}), (\ref{eq:11}) and
(\ref{eq:inertia1}) to be
\begin{equation}
\delta_{\gamma,{\rm osc}}\omega\simeq 
{A_{\rm II}}\omega\,{\rm e}^{-2{\Delta_{\rm II}}^2\omega^2}
\cos2({\tau_{\rm II}\omega}+\epsilon_{\rm II})\,,
\label{eq:14}
\end{equation}
in which we have replaced the phase $\epsilon$ by $\epsilon_{\rm II}$,
and we have introduced a normalized glitch amplitude
\begin{equation}
   A_{\rm II}=\frac{1}{2}\Gamma_{\rm II} T^{-1}\,.
\label{eq:AII}
\end{equation}
\begin{figure}
\centering
\includegraphics[width=1.0\linewidth]{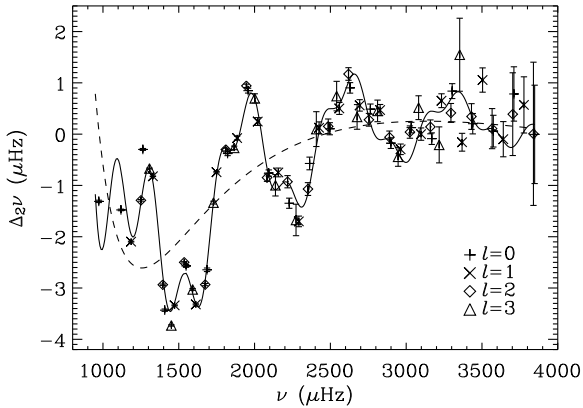}
\caption{
The symbols are second differences $\Delta_2\nu$, defined by 
equation~(\ref{eq:secdiff}), of 
low-degree solar frequencies obtained from GOLF data and kindly supplied
by J. Ballot. Most of those data have been presented by Ballot et al. (2004). 
The most extreme outlying datum has been removed.
The vertical bars represent standard data errors, evaluated
under the assumption that the errors in the raw frequencies are 
independent; their effective overall value is 
$\langle\sigma\rangle=4.3\,$nHz. The curve is the seismic 
diagnostic $D_0(\nu;\alpha_k)$, which has been fitted to the data in a
manner intended to provide an optimal estimate of the eleven parameters 
$\alpha_k$. The values of some of 
those fitting parameters are: $-\delta\gamma/\gamma\vert_{\tau_{\rm II}}=
A_{\rm II}/\sqrt{2\pi}\nu_0\Delta_{\rm II}\simeq0.065$, 
$\tau_{\rm II}\simeq707\,$s, $\Delta_{\rm II}\simeq52\,$s. 
The measure $E$ of the overall misfit is 50\,nHz. The direct measure 
$\overline\chi$ is 9.9; the minimum-$\chi^2$ fit of the function $D_0$ to 
the data yields $\overline\chi_{\rm min}=7.0$.
}
\label{fig:GOLF}
\end{figure}
Notice that the amplitude of the oscillatory component
$\delta_{\gamma,{\rm osc}}\omega$ decays exponentially with
$\omega^2$, which for large $\omega^2$ is faster than the algebraic
decay predicted by the representations of glitches containing
discontinuities (cf. equation~(\ref{eq:16})); the amplitudes due to
those perturbations decay only as $\omega^{-q}$, where $q$ is the order of
the lowest derivative of a dynamically relevant variable of the
equilibrium state that contains a discontinuity.

The acoustic glitch at the base of the convection zone $(r=r_{\rm c})$
also contributes an oscillatory component, $\delta_{\rm c,osc}\omega$, to
the frequency $\omega$ of the modes, and hence to
$\Delta_2\nu$. This component must be included when fitting functional
forms for $\Delta_2\nu$ to real data, even though, in this investigation,
our interest is restricted to helium ionization. The glitch is
essentially a discontinuity in the acoustic cutoff frequency, and is
reflected as a near discontinuity in $\partial\xi/\partial r$, but
not in $\xi$ (a local mixing-length model -- as we have adopted in
the construction of our theoretical test models -- leads to a true
discontinuity). From Appendix\,\ref{sec:delomc} we derive the 
following approximate expression (see equation\,(\ref{eq:delomc})):
\begin{eqnarray}
\delta_{\rm c,osc}\omega&\simeq&
      {A_{\rm c}}\omega^3_0\omega^{-2}
      \left(1+1/4\tau_0^2\omega^2\right)^{-1/2}\cr
      &&\hspace{-10pt}
      \times\,\cos\left[2({\tau_{\rm c}\omega}+\epsilon_{\rm c})
      +\tan^{-1}(2\tau_{\rm 0}\omega)\right]\,,
\label{eq:16}
\end{eqnarray}
with
\begin{equation}
{A_{\rm c}}=\frac{c^2_{\rm c}}{8\pi\omega^2_0}
\left[\frac{{\rm d}^2\ln\rho}{{\rm d}r^2} \right]
^{r_{\rm c}{+}}_{r_{\rm c}{-}} \;\;\,,
\label{eq:ac}
\end{equation}
where $\tau_{\rm c}$ is the acoustic depth of the base of the
convection zone and $c_{\rm c}=c(r_{\rm c})$; $\tau_0\ll T$ is a
measure of the characteristic acoustic distance in the radiative
interior over which the glitch causes the acoustic cutoff frequency to
deviate substantially from a smooth function, and was determined to be
80\,s by fitting an exponential function to a theoretical reference solar
model (see Appendix\,\ref{sec:delomc}). The (constant)
quantity $\omega_0$ is the asymptotic mean large frequency separation:
\begin{equation}
\omega_0=\left(\frac{1}{\pi}\int_0^Rc^{-1}\,{\rm d}r\right)^{-1}
        =\frac{\pi}{T}\,,
\label{eq:nu0}
\end{equation}
which was determined from fitting by least-squares
the asymptotic
expression for the oscillation frequencies \cite[e.g.][]{tass80, g86b}
to the frequency data, namely
\begin{equation}
\nu_{n,l}\sim(n+{\textstyle\frac{1}{2}}\,l+\varepsilon)\nu_0 -
\frac{B\,l(l+1)-C}{\nu_{n,l}}\,\nu^2_0\,,  
\label{eq:new19}
\end{equation}
in which $\nu_0=\omega_0/2\pi$ and $\varepsilon,~B$ and $C$ are also 
constants. The amplitude $A_{\rm c}$ is a measure of the discontinuity 
in the second derivative
of $\rho$ at the base of the convection zone. The amplitude factor
$(1+1/4\tau^2_0\omega^2)^{-1/2}$
is only a weak function of $\omega$, especially above cyclic 
frequencies of about 3~mHz in the solar case.

The amplitudes $A_{\rm II}$ and $A_{\rm c}$, and the phase increments
$2\epsilon_{\rm II}$ and $2\epsilon_{\rm c}+\tan^{-1}(2\tau_0\omega)$, vary 
only slowly with $\omega$. This property facilitates the 
evaluation of the second differences of the oscillatory frequency 
perturbations given by equations (\ref{eq:14}) and (\ref{eq:16}), 
which is described in Appendix\,\ref{sec:DOGcallsthisAppxC}, in 
which $\epsilon_{\rm II}$ and $\epsilon_{\rm c}$ are taken to be
constants. Those constants cannot be taken to be the same because
between $\tau_{\rm II}$ and $\tau_{\rm c}$ the relation
$K=\omega/c$ is not exact. Indeed, this property also provides
part of the reason why these constants cannot be written in 
terms of the phase $\varepsilon$ appearing in the asymptotic
eigenfrequency equation\,(\ref{eq:new19}).
Smooth contributions to the
second differences of the eigenfrequencies (arising, in part, from
refraction in the stellar core, from hydrogen ionization and from the
superadiabaticity of the upper boundary layer of the convection zone
which lies in an evanescent region of the acoustic mode) must also be
accounted for. Guided by the functional form of the asymptotic
relation\,(\ref{eq:new19}), we approximate them
by a third-degree polynomial in $\omega^{-1}$. They
account for both the second frequency differences of a putative
reference solar model that has no glitches associated with helium
ionization or the base of the convection zone, and the smooth
contributions from those glitches. The complete fitting function is then
given by
\begin{equation}
{\Delta_2\nu}\simeq
    \Delta_2\left(\delta_{\gamma,{\rm osc}}\nu+\delta_{\rm c,osc}\nu\right)
                  +\sum_{k=0}^3{a_k\nu^{-k}}\\
\label{eq:sd1a}
\end{equation}
\begin{eqnarray}
&\simeq&F_{\rm II}A_{\rm II}\nu\,{\rm e}^{-8\pi^2\Delta_{\rm II}^2\nu^2}
  \cos[2(2\pi\tau_{\rm II}\nu +\epsilon_{\rm II})-\delta_{\rm II}] \nonumber\\
&&\hspace{-10pt}
+\,F_{\rm c}A_{\rm c}\nu_0^3\nu^{-2}
  \left(1+1/16\pi^2\tau_0^2\nu^2\right)^{-1/2}\cr
&&
\times\,\cos[2(2\pi\tau_{\rm c}\nu+\epsilon_{\rm c})+\tan^{-1}(4\pi\tau_0\nu)-\delta_{\rm c}]
   \nonumber\\
&&\hspace{-10pt}
+\,\sum^3_{k=0}a_k\nu^{-k}\equiv D_0(\nu;\alpha_k)\,,
\label{eq:22}
\end{eqnarray}
where we have now converted to cyclic frequency $\nu$, which we regard
as a continuous variable; formulae for the amplitude factors $F_{\rm II}$ 
and $F_{\rm c}$ and for the phase increments $\delta_{\rm II}$ and 
$\delta_{\rm c}$ arising from taking the second differences are given 
by equations (\ref{eq:DOG_F}) and (\ref{eq:DOG_del})
with $a$ and $b$ given immediately following equation (\ref{eq:ab_II}) in
the case of $F_{\rm II}$ and $\delta_{\rm II}$ and by equations
(\ref{eq:ab_c}) in the case of $F_{\rm c}$ and $\delta_{\rm c}$.
The quantities $\alpha_k$ are the eleven parameters $A_{\rm II},
\Delta_{\rm II}, \tau_{\rm II}, \epsilon_{\rm II}, A_{\rm c}, 
\tau_{\rm c}, \epsilon_{\rm c}$ and $a_k (k=0, ..., 3)$.

In Fig.~\ref{fig:GOLF}, second differences $\Delta_2\nu$ of solar
oscillation frequencies obtained from the Global Oscillations at Low
Frequency (GOLF) investigation
on the Solar and Heliospheric Observatory (SOHO) spacecraft are
plotted against $\nu$. For comparison, the seismic diagnostic $D_0$
is plotted as a continuous curve against $\nu$, which has been fitted
to the data with the intention of providing an optimal estimate of the
eleven parameters $\alpha_k$ by minimizing
$E^2=({\boldsymbol D}_0-{\boldsymbol d})^{\rm t}{\rm C}^{-1}
({\boldsymbol D}_0-{\boldsymbol d})/\sum_k\lambda^{-1}_k$, where 
$\boldsymbol d$ is the vector whose components are
the $N$ data $\Delta_2\nu_{n,l}$, ${\boldsymbol D}_0$ is the $N$-dimensional 
vector with components $D_0(\nu_{n,l};\alpha_k)$, $\rm C$ is the covariance 
matrix of the errors in the data $\boldsymbol d$, and the superscript 
${\rm t}$ denotes transpose. The covariance matrix $\rm C$, whose
eigenvalues are denoted by $\lambda_k$, was evaluated under the 
assumption that the errors in the measurements of the frequencies $\nu_{n,l}$
are independent random variables with variance $\sigma^2_{n,l}$; we 
take the square root of the harmonic mean of $\lambda_k$ as an 
overall measure $\langle\sigma\rangle$ of the magnitude of the 
errors in the second differences. A direct measure of the fit 
of the curve to the data is given by 
$\overline\chi^2=N^{-1}
\sum_{n,l}\{[D_0(\nu_{n,l};\alpha_k)-\Delta_2\nu_{n,l}]/\sigma_{n,l}\}^2$.

A similar comparison may be made with comparable second differences
of frequencies obtained from the Michelson Doppler Imager (MDI), also
on SOHO, which, over the available frequency range, 1473--3711$\,\mu$Hz,
of those data (from which, for equity, we have also removed the most extreme
outlier) have overall error $\langle\sigma\rangle=5.3\,$nHz,
and deviate from the fitted diagnostic~(\ref{eq:22}) by $E=32\,$nHz,
compared with $\langle\sigma\rangle=6.4\,$nHz and $E=34\,$nHz
for the GOLF data over the same frequency range. The sharp upturn of 
the purported `smooth'
contribution (dashed curve) in Fig.~\ref{fig:GOLF} as $\nu$ decreases 
below 1300$\,\mu$Hz is probably an artefact of the failure of the 
asymptotic approximation upon which equation~(\ref{eq:22}) is founded; 
this deficiency is removed partially by the improved diagnostic $D_2$ 
which we introduce in Section\,\ref{sec:fimprov}.

One should expect equation (\ref{eq:22}) to represent the second
differences of the actual frequencies more faithfully than the
expressions that have been used before
\citep[e.g.,][]{mt98,mt05,g02,b04}. This is so partly because here we
use a representation of the \HeII\ glitch in $\gamma$ that is more
realistic than the non-analytic functions adopted by Monteiro\,et\,al.
and \citeauthor{b04}, and we also use 
a more realistic representation of the
stratification immediately beneath the convection zone.

\begin{figure}
\centering
\includegraphics[width=0.60\linewidth]{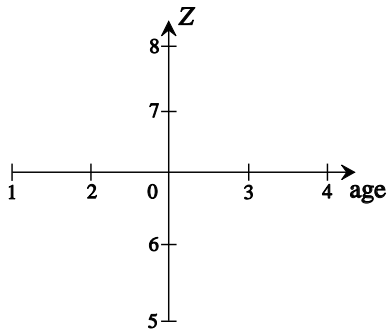}
\caption{
Denotation of the nine calibrated solar models which are used for testing 
the formulation of the second frequency differences. The `central model' 
is model~0; the set of the four models (1-4) have a constant value of $Z$ 
but varying age; the second set of the models (5-8) have constant age but 
varying $Z$.} 
\label{fig:nine-models}
\end{figure}


   \section{Testing the formulation on theoretical models}
   \label{sec:testing}


\begin{figure}
\centering
\includegraphics[width=1.0\linewidth]{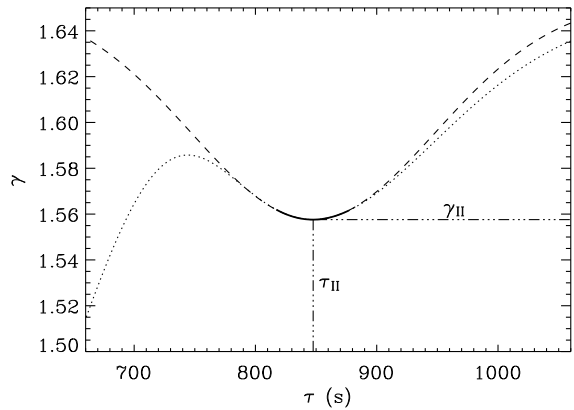}
\caption{
Adiabatic exponent $\gamma$ (solid and dotted curve) for the central 
stellar model\,0 as a function of the acoustic depth $\tau$, which is
measured from the acoustic surface of the model (estimated as the place
where the linear extrapolation of $c^2$ from the region of the upper turning
points of the modes vanishes, some 225\,s above the photosphere).
The dashed curve is $\gamma_{\rm II0}- (\gamma_{\rm II0}-\gamma_{\rm II})
\exp[-(\tau-\tau_{\rm II})^2/2\Delta_{\rm II}^2]$,
where $\gamma_{\rm II0}=1.651$ is the value of 
$\gamma(p(\tau_{\rm II}), \vartheta(\tau_{\rm II}), 0)$, evaluated at $Y=0$
(dot-dashed curve in Fig.~\ref{fig:gamma_taylor}) and $\vartheta$ is 
temperature; the remaining 
parameters ($\gamma_{\rm II}$, $\tau_{\rm II}$, $\Delta_{\rm II}$) have 
been adjusted to fit $\gamma$ in the region in which the curve 
representing $\gamma$ is solid, namely 
$\tau_{\rm II}-\tau_{\rm f}<\tau<\tau_{\rm II}+\tau_{\rm f}$
with $\tau_{\rm f}\simeq30\,$s.
The parameter $\gamma_{\rm II}$ is the minimum value of $\gamma$ in 
the \HeII\ ionization zone, and $\tau_{\rm II}$ is the acoustic depth at 
that minimum.
The parameters associated with the other test models were determined 
likewise, with values of $\tau_{\rm f}$ as close as possible to that of the
central model subject to their fitting exactly onto the computational mesh;
their values differ from that of the central model by less than 3\,s.
} 
\label{fig:gamma_fit}
\end{figure}


\subsection{Test models}
\label{sec:testmodels}

Two sets of five calibrated evolutionary models for the Sun were used. 
The models had been computed by \citet{gn90} using the procedure described
by \citet{cd81}. The opacity values were obtained from interpolating in
the Los Alamos Opacity Library for the \citet{gr84} mixture, 
supplemented at low temperature by interpolating in the tables 
of \citet{cs70a,cs70b}. Gravitational settling was not included.
One set of models has a constant value for the heavy-element abundance $Z$ 
but varying age in (logarithmically) uniform steps of magnitude 
approximately 0.051 in $\ln t$ ($t$ being age); the other 
has constant age but 
varying $Z$, in uniform steps of magnitude approximately 0.016 in 
$\ln Z$. The central models (model 0, see Fig.~\ref{fig:nine-models})
of the two sets coincide, having $t=4.60\,$Gy and $Z=0.02014$. All models
were calibrated by adjusting the initial helium abundance $Y_0$ and the
mixing-length parameter to satisfy the solar values of the luminosity and
photospheric radius: 
${\rm L}_\odot=3.845\times10^{33}\,$erg\,s$^{-1}$ and 
${\rm R}_\odot=6.9626\times10^{10}\,$cm, resulting in a central model
with $Y_0=0.2762$.
The models were examined carefully and 
adjusted to satisfy hydrostatic equilibrium to high precision. 
Eigenfrequencies of adiabatic oscillation modes with $l\le2$ were 
computed for all models in the Cowling approximation
(gravitational forces are long-range and do not materially affect glitch
diagnostics, so the Cowling approximation is quite adequate for our purposes)
using the {\it Aarhus Adiabatic Pulsation Package}\footnote{
http://astro.phys.au.dk/$\sim$jcd/adipack.n}. From these were constructed
second frequency differences according to equation\,(\ref{eq:secdiff}).

\subsection{Test results}
\label{sec:test}

\begin{figure*}
\centering
\includegraphics[angle=90,width=0.76\linewidth]{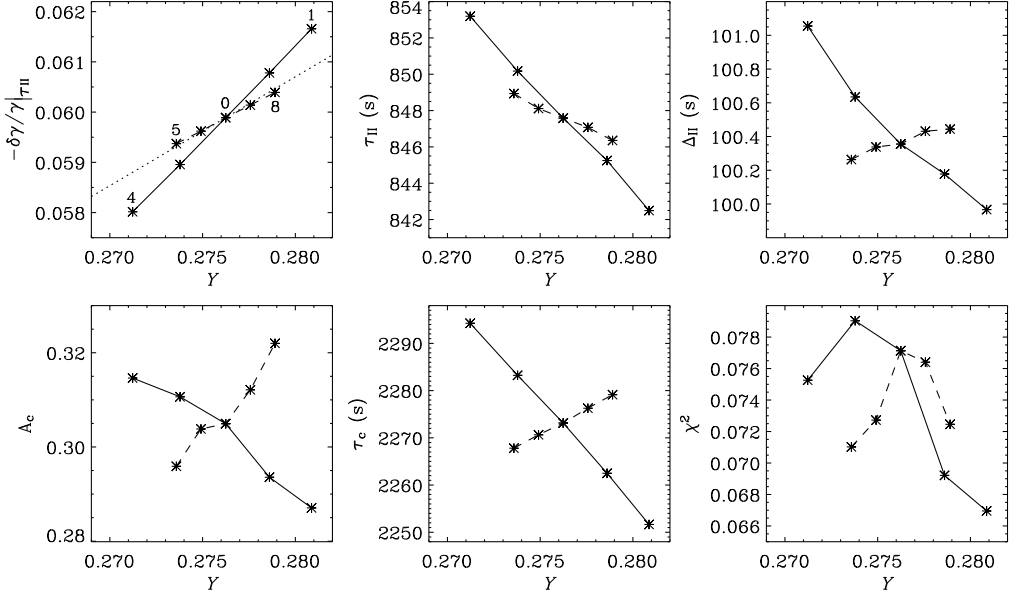}
\caption{
The top panels show $-\delta\gamma/\gamma\vert_{\tau_{\rm II}}:=
2(\gamma_{\rm II0}-\gamma_{\rm II})/(\gamma_{\rm II0}+\gamma_{\rm II})$, 
$\tau_{\rm II}$ and $\Delta_{\rm II}$, which were determined from fitting 
a Gaussian (see Fig.~\ref{fig:gamma_fit}) to the \HeII\ depression in $\gamma$, 
for all nine test models, plotted against helium abundance $Y$.
Models with varying age and constant heavy element abundance $Z$ are connected 
with solid lines; models with varying $Z$ and constant age are connected with 
dashed lines. The central model 0 and the models 1, 4, 5 and 8 are indicated 
in the upper left panel. Also in the upper left panel is a dotted straight
line from the origin through the central model; it is evident that the
effective amplitude of the helium-induced acoustic glitch is consistent
with it being a linear function of $Y$ in the $Z$-varying sequence, but 
not strictly so in the sequence with varying age.
The lower right panel is the standard 
measure~$\chi^2$ (normalized integral of squared residuals:
$\chi^2=(2\Delta_{\rm II})^{-1}\int(\delta\gamma/\gamma-G)^2\,{\rm d}\tau$, in
which the integral is evaluated from $\tau_{\rm II}-\Delta_{\rm II}$ to
$\tau_{\rm II}+\Delta_{\rm II}$) 
of the goodness of the fit between $\gamma$ and the calibrated Gaussian.
The first two lower panels show the properties of the density discontinuity 
(amplitude $A_{\rm c}$ and acoustic depth $\tau_{\rm c}$) at the base of the convection 
zone, which were determined from the equilibrium models.
}
\label{fig:gamma_nine-fit}
\end{figure*}

For all nine test models we determined the properties of the \HeII\ depression 
in $\gamma$ by fitting a Gaussian (plus a constant) to $\gamma$. Details of 
this fitting process are illustrated in Fig.~\ref{fig:gamma_fit} for the 
central model 0.
The three parameters, $-\delta\gamma/\gamma\vert_{\tau_{\rm II}}$, 
$\tau_{\rm II}$ and $\Delta_{\rm II}$ so determined are plotted in 
the top three panels of Fig.~\ref{fig:gamma_nine-fit} for the nine test models 
as a function of the helium abundance $Y$, where 
$-\delta\gamma/\gamma\vert_{\tau_{\rm II}}
\equiv2(\gamma_{\rm II0}-\gamma_{\rm II})/(\gamma_{\rm II0}+\gamma_{\rm II})$.
The value of $\gamma_{\rm II0}$ is the value of $\gamma$ evaluated with the
pressure and temperature of the model at $\tau=\tau_{\rm II}$, but with 
a helium abundance $Y=0$ (see dotted curve in Fig.~\ref{fig:gamma_taylor}).
For the central model $\gamma_{\rm II0}=1.651$. The values of the other models
differ from that of the central model by less than 2$\times10^{-4}$.


\begin{figure*}
\centering
\includegraphics[angle=90,width=0.76\linewidth]{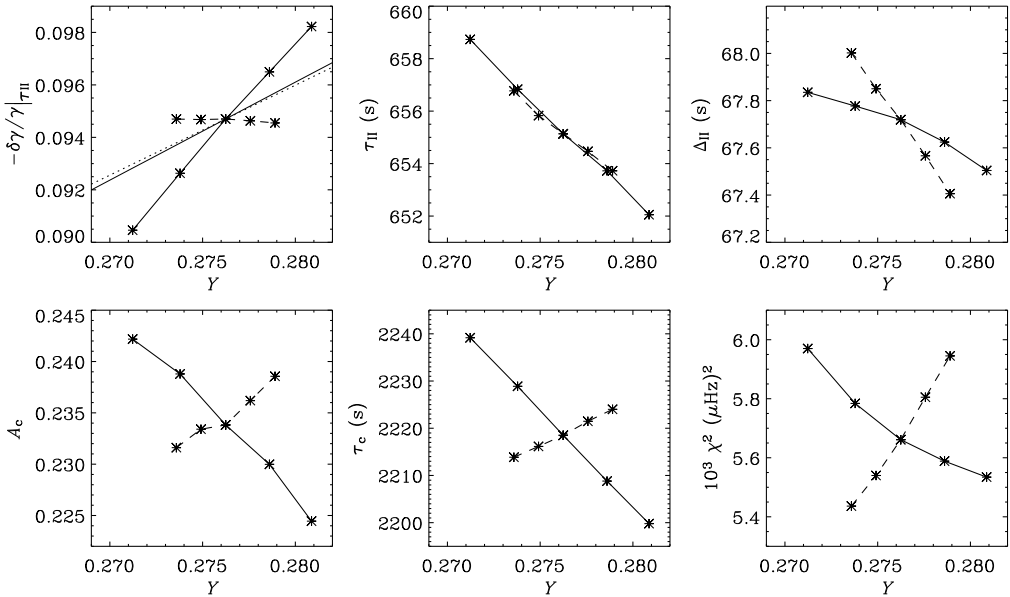}
\caption{Model properties 
$-\delta\gamma/\gamma\vert_{\tau_{\rm II}}=
A_{\rm II}/\sqrt{2\pi}\nu_0\Delta_{\rm II}$, 
$\tau_{\rm II}$, $\Delta_{\rm II}$, $A_{\rm c}$ and $\tau_{\rm c}$ for all 
nine test models determined from the diagnostic $D_0$
defined by equation (\ref{eq:22}).
The fitting parameters $A_{\rm II}$, $\tau_{\rm II}$, $\Delta_{\rm II}$, 
$\epsilon_{\rm II}$, $A_{\rm c}$, $\tau_{\rm c}$ and $\epsilon_{\rm c}$ have 
been adjusted to fit the second differences $\Delta_2\nu$, defined by 
equation~(\ref{eq:secdiff}), of the low-degree ($l$=0,1,2), adiabatically 
computed, model eigenfrequencies. Note that the inferred values of 
$-\delta\gamma/\gamma\vert_{\tau_{\rm II}}$ actually decrease, gradually, with 
increasing $Y$ in the $Z$-varying sequence. The frequency range used in the 
least-squares fitting is that available to the BiSON data, with which we 
compare our final diagnostic formula in Fig.~\ref{fig:fit_BiSON}: it is 
1322--4058$\,\mu{\rm Hz}$. The lower right panel is the standard 
measure~$\chi^2$ (mean squared residuals, in units of $(\mu{\rm Hz})^2$)
of the goodness of the fit between $\Delta_2\nu$ of the 
modelled frequencies and the calibrated diagnostic.
The line styles are as in Fig.~\ref{fig:gamma_nine-fit}.
The dotted line in the upper left panel is again a straight line from 
the origin; for comparison the thin solid line is drawn with the 
same slope as the age-varying sequence in Fig.~\ref{fig:gamma_nine-fit}.
}
\label{fig:sd1ab}
\end{figure*}
\begin{figure*}
\centering
\includegraphics[angle=90,width=0.76\linewidth]{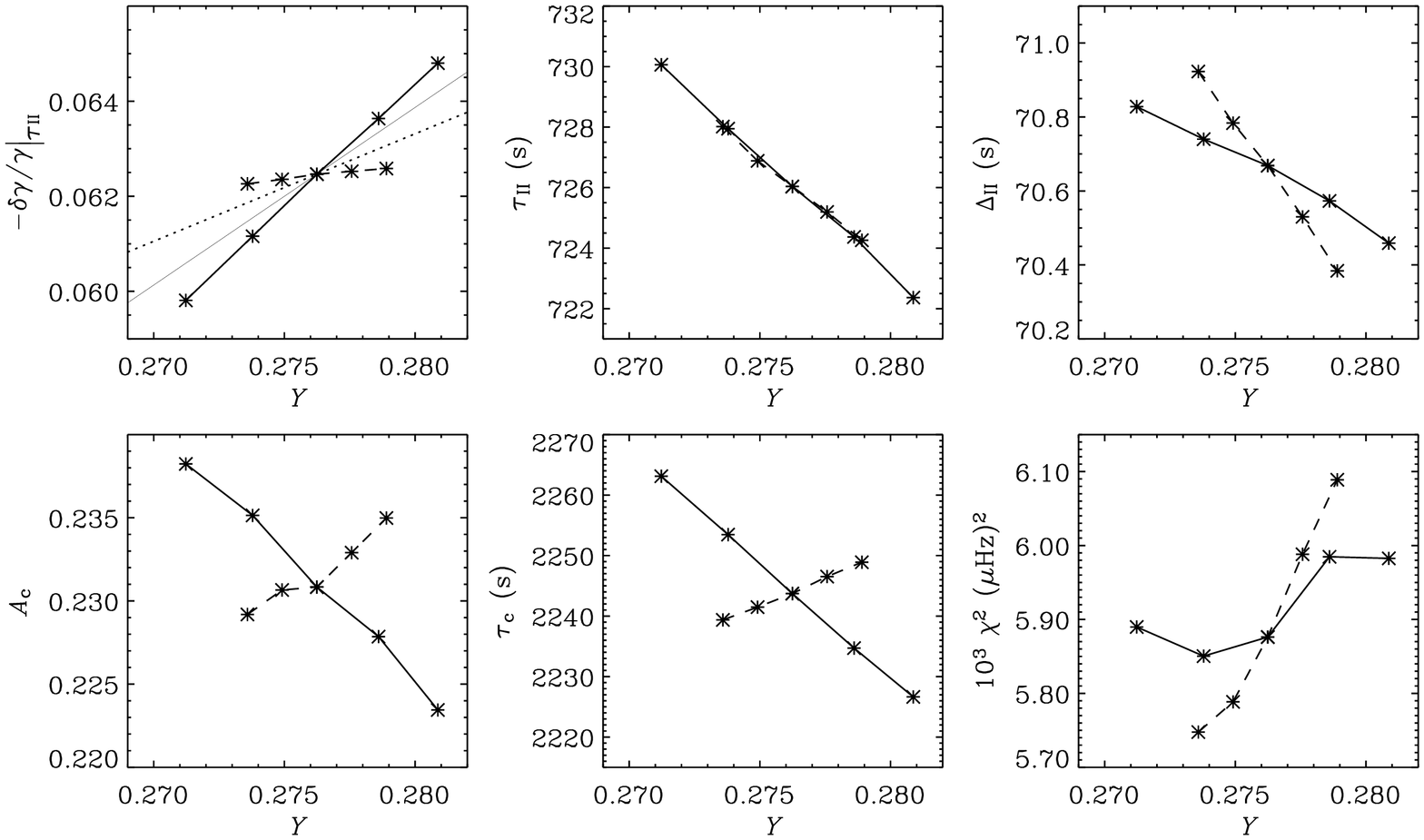}
\caption{Model properties $-\delta\gamma/\gamma\vert_{\tau_{\rm II}}$, 
$\tau_{\rm II}$, $\Delta_{\rm II}$, $A_{\rm c}$ and $\tau_{\rm c}$ determined
from the diagnostic $D_1$ defined by equations (\ref{eq:sd1a}), 
(\ref{eq:28}) \& (\ref{eq:29}) 
which includes an improved treatment of the phase 
function $\psi$ given by equation~(\ref{eq:phi1}).
Results are shown for the nine test models. 
The fitting parameters $A_{\rm II}$, $\tau_{\rm II}$, $\Delta_{\rm II}$, 
$\epsilon_{\rm II}$, $A_{\rm c}$, $\tau_{\rm c}$ and $\epsilon_{\rm c}$ have 
been adjusted to fit 
the second differences $\Delta_2\nu$, defined by equation (\ref{eq:secdiff}), 
of the low-degree ($l$=0,1,2), adiabatically computed, model frequencies. The 
smallest frequency value used in the least-squares fitting is 
$\nu\simeq1322\,\mu{\rm Hz}$, the largest is $4058\,\mu{\rm Hz}$. 
The lower right panel is the standard measure~$\chi^2$ 
(mean squared residuals, in units of $(\mu{\rm Hz})^2$)
of the goodness of the fit between $\Delta_2\nu$ of the 
modelled frequencies and the calibrated diagnostic.
The line styles are as in Fig.~\ref{fig:gamma_nine-fit}.
The dotted line in the upper left panel is a straight line from 
the origin; the thin solid line is drawn with the 
same slope as the age-varying sequence in Fig.~\ref{fig:gamma_nine-fit}.
}
\label{fig:sd1a_delphi}
\end{figure*}
The first two lower panels in Fig.~\ref{fig:gamma_nine-fit} display the
value of $A_{\rm c}$ (equation\,(\ref{eq:ac})) and the acoustic depth 
$\tau_{\rm c}$ of the base of the convection zone. 
These values were obtained by fitting analytical functions (as discussed by
\citealt{cgt91}) to the first logarithmic density derivative, 
${\rm d}\ln\rho/{\rm d}\ln r$,
of the models either side of the discontinuity at the base of the
convection zone, as explained in Appendix\,\ref{sec:appendixD}.
The third (rightmost) panel is a standard measure (mean squared residuals) 
$\chi^2$ for the goodness of the Gaussian fit to the \HeII\ depression in 
$\gamma$. 

The asymptotic formula $D_0$ was then fitted to the second
differences of the computed eigenfrequencies of the models as
described in Section~\ref{sec:seismodel}; some of the resulting
parameters are plotted in Fig.~\ref{fig:sd1ab}. There is a superficial
resemblance to the corresponding direct estimates from the models
depicted in Fig.~\ref{fig:gamma_nine-fit} (except for the slope of
$\Delta_{\rm II}(Y)$ for the $Z$-varying sequence, which we suspect is
influenced substantially by heavy-element ionization, and, of course,
$\chi^2$, which has quite different meanings in the two figures). 
Perhaps the most obvious other difference is that the slopes of the
$\delta\gamma/\gamma\vert_{\tau_{\rm II}}$ curves from the two sequences
of direct model measurements are almost the same, whereas those of the
frequency-fitting are not. The magnitudes of the values of
some of the parameters are also discrepant. As will become
evident in Section\,\ref{sec:fimprov}, that is to be expected in the case 
of the parameters characterizing the ionization, partly because
the seismological fit is influenced by the effect of \HeI\ ionization,
which has been treated erroneously as being part of \HeII. The
acoustic depths of the bases of the convection zones agree reasonably
well. But the amplitudes do not; we do not know why.

  \section{Further improvements}
  \label{sec:fimprov}

\subsection{Improved treatment of the asymptotic phase function $\bpsi$}
\label{sec:iphase}

\noindent One of the obvious deficiencies of our introductory analysis
described in the previous section is our treatment of the phase
function $\psi$, which we approximated by $\psi\simeq\omega\tau+\epsilon$,
neglecting the acoustic cutoff frequency
$\omega_{\rm a}$ (and also the relatively small horizontal component
of the wavenumber) in evaluating the vertical component $K$ of the wavenumber,
whose full planar form is given by equation~(\ref{eq:B2}).
In this section we account for the effect of the frequency dependence of the
upper turning point by keeping $\omega_{\rm a}$:
\begin{equation}
K(r)\simeq{\omega\over c} \left(1-{\omega^2_{\rm
a}\over\omega^2}\right)^{1/2}\,,
\end{equation}
representing the stratification of the outer stellar layers
locally by a polytrope of index $m$, for which the acoustic cutoff
frequency associated with the Lagrangian pressure perturbation (and
corrected for the singularity at the top of the polytropic envelope --
see~\citealt{g93}) is
\begin{equation}
\omega_{\rm a}=(m+1)/\tau\,.
\label{eq:omega_a}
\end{equation}
Thus, recognizing that the acoustic glitch associated with the helium
ionization is confined to a small range of acoustic depth, we expand
the phase function $\psi(\tau)$ about the (phase-shifted) centre of the 
glitch, $\tau=\tilde\tau_{\rm II}$, having accommodated the phase constant 
$\epsilon_{\rm II}$ by replacing $\omega\tau_{\rm II}$ by 
$\omega\tilde{\tau}_{\rm II}=\omega\tau_{\rm II}+\epsilon_{\rm II}$, yielding
\begin{equation}
\psi(\tau)\simeq\psi(\tilde\tau_{\rm II})
+\omega \kappa_{\rm II}(\tau-\tilde\tau_{\rm II})
+{\rm O}\!\left[(\tau-\tilde\tau_{\rm II})^2\right]\,,
\label{eq:phiexp}
\end{equation}
with 
$\kappa_{\rm II}=\kappa(\tilde{\tau}_{\rm II})$, where
\begin{equation}
\kappa(\tau)\equiv\sqrt{1-\left({m+1\over\omega\tau}\right)^2}\,.
\label{eq:phi2}
\end{equation}
We adopt $m=3.5$ for all models; it was obtained from the slope of
$1/\omega_{\rm a}(\tau)$ in the outer layers of the adiabatically stratified
region of the convection zone in the central model, and relating that 
to $m$ using equation\,(\ref{eq:omega_a}).
We note, in passing, that this value is close to 3.0, the effective 
value of $m$ in the region of the upper turning points of solar modes
of intermediate degree determined by calibrating the asymptotic eigenfrequency
formula~(\ref{eq:asympfrequ}) against observation \citep[e.g.][]{cdghr85}; 
actually, the values of $\chi^2$ of the data-fits that follow are reduced 
somewhat if the value 3.5 is replaced by 3.0, but then the overestimates of 
corresponding values of the amplitudes 
$-\delta\gamma/\gamma\vert_{\tau_{\rm II}}$ are greater.
Taking expansion (\ref{eq:phiexp}) to higher order makes a correction of
magnitude typically less than one per cent, so we keep only these
first two terms. The leading term in the expansion (\ref{eq:phiexp})
is given by
\begin{eqnarray}
\psi(\tilde\tau_{\rm II})&\!\!\!\simeq\!\!\!&\omega
                                \int^{(m+1)/\omega}_{\tilde\tau_{\rm II}}
\kappa\,{\rm d}\tau+\frac{\pi}{4}\cr 
&\!\!\!=\!\!\!&\kappa_{\rm II}\tilde{\tau}_{\rm II}\omega-(m+1)\cos^{-1}
\left(\frac{m+1}{\tilde{\tau}_{\rm II}\omega}\right)+\frac{\pi}{4}~.
\label{eq:phi1}
\end{eqnarray}
The analysis proceeds as in Section\,\ref{sec:seismodel} and 
Appendix\,\ref{sec:varprin}. 
The outcome is to change the inertia $\cal{I}$ to 
\begin{eqnarray}
{\cal{I}}&\simeq&\psi(T)-\frac{\pi}{4}
\;\simeq\;\frac{1}{2}T\omega-\frac{1}{4}(m+1)\pi\cr
&\simeq&\frac{\pi\nu^2}{2\nu_0}\left[\nu+{\textstyle\frac{1}{2}}(m+1)\nu_0\right]^{-1}\;.
\label{eq:inertia2}
\end{eqnarray}
The contribution to $\nu$ from the glitch then becomes
\begin{eqnarray}
\delta_\gamma\nu\simeq 
A_{\rm II}\kappa_{\rm II}^{-1}\left[\nu+{\textstyle\frac{1}{2}}(m+1)\nu_0\right]\!
    &&\hspace{-20pt}
\left[{\rm e}^{-8\pi^2\kappa^2_{\rm II}\Delta^2_{\rm II}\nu^2}\!\cos(2\psi_{\rm II})\!-\!1\right],\cr
&&
\label{eq:28}
\end{eqnarray}
in which $A_{\rm II}$ is still given by equation (\ref{eq:AII}),
and $\psi_{\rm II}=\psi(\tilde{\tau}_{\rm II})$, with 
$\psi(\tilde{\tau}_{\rm II})$ given by equation (\ref{eq:phi1}).
The effect on the oscillatory component $\delta_{\gamma,{\rm osc}}\nu$ is 
both to modify the amplitude of the cosine in equation 
(\ref{eq:14}) and effectively to impart an additional
frequency-dependent contribution to the
phase.

Beneath the convection zone the $\omega_{\rm a}$ correction is small,
and can be ignored; for the Sun, for example,
$1-\kappa(\tau)\,<\,$\hbox{$1\times10^{-2}$} for $\tau\ge\tau_{\rm c}$. 
Moreover, the measure ${\cal I}_\Psi$ of the inertia, introduced in 
Appendix\,\ref{sec:delomc}, is unchanged.
Consequently the contribution from the acoustic glitch at
the base of the convection zone is influenced significantly by 
$\omega_{\rm a}$ only through the phase 
$\psi_{\rm c}=\psi(\tilde{\tau}_{\rm c})$, where
$\omega\tilde{\tau}_{\rm c} =\omega\tau +\epsilon_{\rm c}$, leading to
\vspace{-3pt}
\begin{eqnarray}
\delta_{\rm c}\nu\!\!&\simeq&\!\!A_{\rm c}\nu_0^3\nu^{-2}
   \left(1+1/16\pi^2\tau_0^2\nu^2\right)^{-1/2}\cr
&\times&\hspace{-8pt}\left\{\cos[2\psi_{\rm c}+\tan^{-1}(4\pi\tau_0\nu)]
      \!-\!(16\pi^2\tau_0^2\nu^2\!+\!1)^{1/2}
\right\},
\label{eq:29}
\end{eqnarray}
where $A_{\rm c}$ is given by equation (\ref{eq:ac}) and
$\psi_{\rm c}$ by equation~(\ref{eq:newB20}).

The improved diagnostic function $D_1(\nu;\alpha_k)$ for representing the 
second frequency differences associated 
with the acoustic glitches is then given by equation~(\ref{eq:22}) with
the first term modified according to the oscillatory component of 
equation~(\ref{eq:28}) and the second term by equation~(\ref{eq:29}), 
and still with $F_{\rm II}$ and $\delta_{\rm II}$ defined by equations
(\ref{eq:DOG_F}) and (\ref{eq:DOG_del}) but with the extended expressions
for $\alpha$, $a$ and $b$ given by equations (\ref{eq:newC7}) and 
(\ref{eq:newC8}). The form of the seismic diagnostic from the base of the
convection zone is unaffected, save for the replacement
of $2\pi\tau_{\rm c}\nu+\epsilon_{\rm c}$ by $\psi_{\rm c}$.\\
Results for the seismic diagnostic $D_1$ are
shown in Fig.~\ref{fig:sd1a_delphi}
for all nine test models (in which we have, as before, considered the 
smooth terms\,--\,i.e. the last terms in both equations (\ref{eq:28}) 
\& (\ref{eq:29})\,--\,to have been incorporated into the polynomial in
$\nu^{-1}$).
Compared with the results of the previous section, plotted in 
Fig.~\ref{fig:sd1ab}, the amplitudes
$-\delta\gamma/\gamma\vert_{\tau_{\rm II}}$ are smaller and the values of
$\tau_{\rm II}$, $\Delta_{\rm II}$ and $\tau_{\rm c}$ are larger, all being 
closer to the model values depicted in Fig.~\ref{fig:gamma_nine-fit}.

\subsection{Including He{\,\sevensize\bf I} ionization}
\label{sec:helI}
\begin{figure}
\centering
\includegraphics[width=1.0\linewidth]{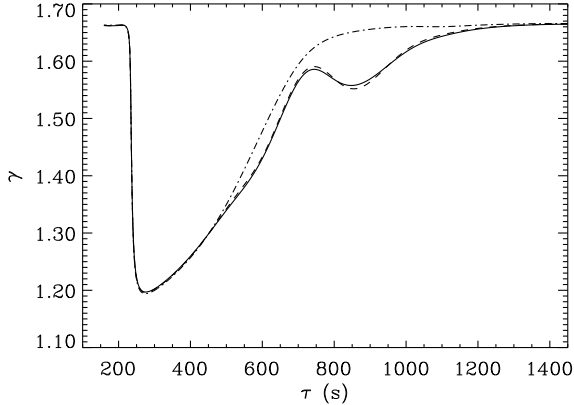}
\caption{ The solid curve is the adiabatic exponent 
$\gamma(p, \vartheta, Y)$, where $\vartheta$ is temperature, through the
ionization zone of the central model\,0; the dot-dashed curve is the
adiabatic exponent $\gamma(p, \vartheta, 0)$. The dashed curve is
the sum of the first two terms of the Taylor expansion of $\gamma$
about $Y=\;$0 given by equation (\ref{eq:gamma_taylor1}). Note that the
amplitude of variation of the dashed curve in the \HeII~ ionization
zone is greater than that of the solid curve; the difference
contributes to the discrepancy between the values of 
$\delta\gamma/\gamma\vert_{\tau_{\rm II}}$ in 
Figs~\ref{fig:gamma2_nine-fit} and \ref{fig:ff3}.}
\label{fig:gamma_taylor}
\end{figure}

A second deficiency in our initial procedure is that we took no explicit
account of the first ionization of helium. The reason is that 
\HeI\ ionization is merged with the ionization of hydrogen, which dominates
in the depression of $\gamma$ even at the centre of \HeI\ ionization. 
However, the effect of the first ionization of helium is, as is evident in 
Fig.~\ref{fig:gamma_taylor}, not negligible; it broadens the
hydrogen-induced dip in $\gamma$. Without precise knowledge
of the hydrogen-induced profile (perhaps even with it) seismological
calibration of the broadening is difficult. Our approach is not even
to try. Instead we simply attempt to relate directly the reduction of
$\gamma$ by \HeI\ ionization to that due to \HeII\ ionization, which
we do aim to measure.
The much greater effect of hydrogen ionization is, on the whole, acoustically
shallower than the effect of \HeI, and therefore produces a smoother seismic
signal, and we trust that it is adequately accommodated by the polynomial 
$\sum a_k\nu^{-k}$\,in the expression\,(\ref{eq:sd1a}) for
$\Delta_2\nu\,$.\\

The degree of helium ionization, and its effect on $\gamma$, is
determined by the law of mass action for ionizing species, which
is approximated by Saha-like equations (we use the approximate equation
of state of \citealt{eff73}). It depends not only on the
temperature and the number density of helium, but also on the
free-electron density, which itself depends on the abundances
of all ionized species, especially hydrogen. In general these
factors are nonlinearly dependent on one another, but we base our
procedure here on the idea that because the number density of helium
is quite low, although not extremely low, the influence on
$\gamma$ is approximately a linear function of the helium
abundance $Y$. As we demonstrate below, the linear approximation
works quite well, although it is evident from the top left panels
of Figs 4, 5, 6, 9 and 10 that the linear approximation is not 
strictly correct. 
It is important to realize, however, that it is used
solely for estimating the \HeI\ contribution to $\gamma$ relative
to the larger \HeII\ contribution, and therefore plays only a minor
role in calibrating $Y$: our expectation is that it can account
adequately for the deviation of the form of the seismic
helium signature from those implied by the simple equations
(\ref{eq:14}) and (\ref{eq:28}), thereby producing a more
faithful match with the data, and consequently permitting a
more robust calibration of the models against the star.\\
We first consider $\gamma$ to be Taylor-expanded about $Y=0$
in a reference stellar model, yielding
\begin{equation}
\gamma(\tau,Y)\simeq\gamma(\tau,0)\big\vert_{Y=0}
            +{\partial\gamma/\partial Y}\big\vert_{Y=0}\,Y\,,
\label{eq:gamma_taylor1}
\end{equation}
where the coefficients $\gamma$ and $\partial\gamma/\partial Y$ on the
right-hand side of equation (\ref{eq:gamma_taylor1}) are evaluated at
the temperature and pressure of the reference model but with $Y=0$. We
hold temperature, $\vartheta$, rather than density, fixed under
differentiation, because ionization depends much more sensitively on
$\vartheta$. We do not take heavy elements explicitly into account
here because $\partial\gamma/\partial Z\vert_{Z=0}$, where $Z$ is the
total heavy element abundance, is rather smaller in magnitude than
$\partial\gamma/\partial Y\vert_{Y=0}$, and in any case $Z$ itself is 
much smaller than $Y$.

\begin{figure}
\centering
\includegraphics[width=1.0\linewidth]{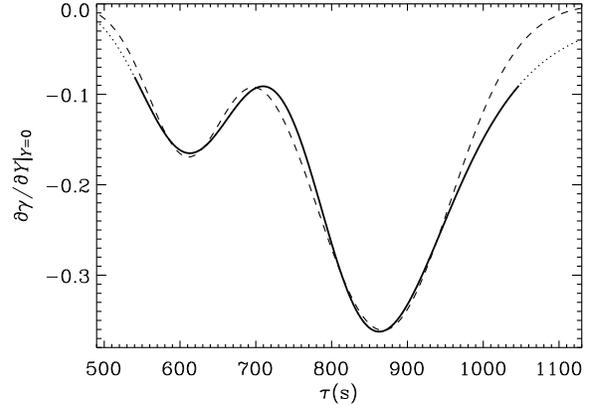}
\caption{ The solid and dotted curve is $\partial\gamma/\partial
Y\vert_{Y=0}$ for the central model~0, the derivative being taken at
constant $p$ and $\vartheta$ (and $Z$). The dashed curve is the sum of the
two Gaussian functions used in the construction of the seismic
diagnostic~(\ref{eq:ff3a}) whose parameters have been adjusted to fit
$\partial\gamma/\partial Y\vert_{Y=0}$ in the region in which the
curve is solid.}
\label{fig:dGdY}
\end{figure}


\begin{figure*}
\centering
\includegraphics[angle=90,width=0.76\linewidth]{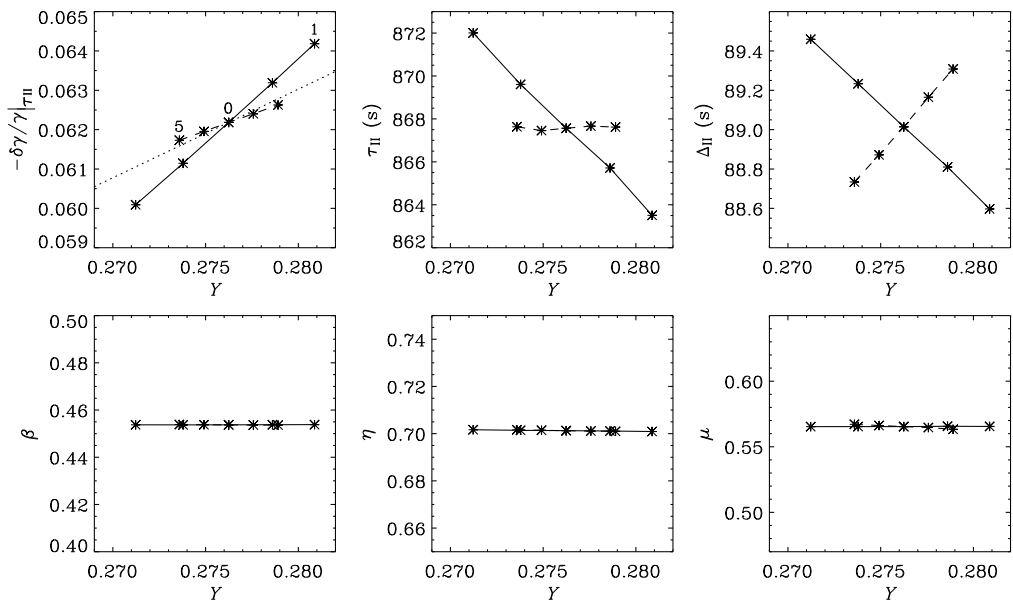}
\caption{
Properties of the \HeII\ depression in $\gamma$
for all nine test models as a function of $Y$.
The \HeII\ properties (top panels), including $\tau_{\rm II}$, 
were obtained from fitting the two Gaussians (equation\,(\ref{eq:2Gau})) to 
$\partial\gamma/\partial Y\vert_{Y=0}$ of the nine test models; 
they are somewhat greater than the acoustic depths of the minima 
in $\gamma$ plotted in Fig.~\ref{fig:gamma_nine-fit}, apparently because
the locations of those minima are influenced by the $\tau$-dependent
contribution from the hydrogen ionization. Partly as a consequence of fitting
the Gaussians over an extended range of $\tau$,
the values of $-\delta\gamma/\gamma\vert_{\tau_{\rm II}}$, which were computed
according to equation~(\ref{eq:gamma_taylor}), are somewhat
greater than those in Fig.~\ref{fig:gamma_nine-fit}, possibly the result of
a contribution from the ionization of the heavy elements; however, the 
difference is not great (cf. Fig.~\ref{fig:gamma_taylor}).
The values (not plotted here) of $A_{\rm c}$ and $\tau_{\rm c}$ are, of course,
the same as those in Fig.~\ref{fig:gamma_nine-fit}.
The line styles are as in Fig.~\ref{fig:gamma_nine-fit}.
Also in the upper left panel is a dotted straight
line from the origin through the central model; the
effective amplitude of the helium-induced acoustic glitch is consistent
with it being a linear function of $Y$ in the $Z$-varying sequence, but 
not strictly so in the sequence with varying age.
The lower panels show the values of $\beta$, $\eta$ and $\mu$.
}
\label{fig:gamma2_nine-fit}
\end{figure*}


\begin{figure*}
\centering
\includegraphics[angle=90,width=0.76\linewidth]{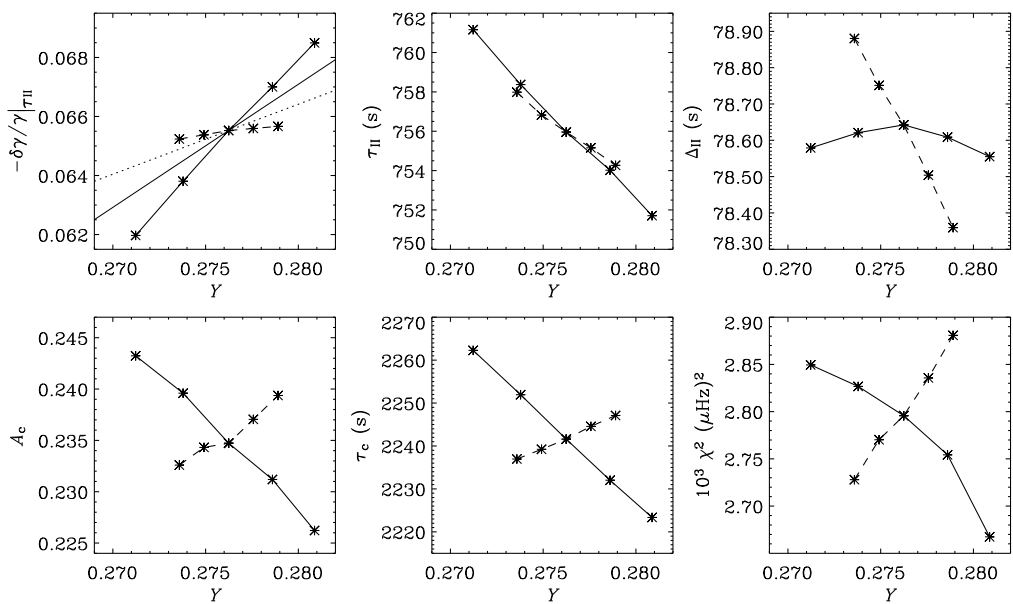}
\caption{
Model properties $-\delta\gamma/\gamma\vert_{\tau_{\rm II}}$, defined as 
in Fig.\,\ref{fig:sd1ab}, $\tau_{\rm II}$, 
$\Delta_{\rm II}$, $A_{\rm c}$ and $\tau_{\rm c}$ determined from the
seismic diagnostic $D_2$ defined by equations (\ref{eq:sd1a}), 
(\ref{eq:29}) \& (\ref{eq:ff3a}), 
which includes a description for both the \HeI\ and the \HeII\ ionization 
zones.  Results are displayed for the nine test models. 
The fitting parameters $A_{\rm II}$, $\tau_{\rm II}$, $\Delta_{\rm II}$, 
$\epsilon_{\rm II}$, $A_{\rm c}$, $\tau_{\rm c}$ and $\epsilon_{\rm c}$ have 
been adjusted to fit 
the second differences $\Delta_2\nu$, defined by equation (\ref{eq:secdiff}), 
of the low-degree ($l$=0,1,2), adiabatically computed, model frequencies. The 
smallest frequency value used in the least-squares fitting is 
$\nu\simeq1322\,\mu{\rm Hz}$, the largest is $4058\,\mu{\rm Hz}$. 
The lower right panel is the standard 
measure~$\chi^2$ of the goodness of the fit between $\Delta_2\nu$ of the 
modelled frequencies and the calibrated seismic diagnostic.
The line styles are as in Fig.~\ref{fig:gamma_nine-fit}.
The dotted line in the upper left panel is a straight line from 
the origin; the thin solid line is drawn with the 
same slope as the age-varying sequence in Fig.~\ref{fig:gamma2_nine-fit}.
}
\label{fig:ff3}
\end{figure*}


The variation $\delta\gamma$ in the adiabatic exponent $\gamma$ 
induced by helium ionization then becomes:
\begin{equation}
\delta\gamma\simeq\partial\gamma/\partial Y\big\vert_{Y=0}\,Y\,. 
\label{eq:gamma_taylor}
\end{equation}
The first-order expansion (\ref{eq:gamma_taylor1}) is illustrated 
in Fig.~\ref{fig:gamma_taylor} for the central model 0;
the solid curve denotes the adiabatic exponent
$\gamma(p, \vartheta, Y)$ of the model, the dotted curve 
is $\gamma(p, \vartheta, 0)$, and the dashed curve is the right-hand side 
of equation (\ref{eq:gamma_taylor1}).
A plot of $\partial\gamma/\partial Y\vert_{Y=0}$ is presented in 
Fig.~\ref{fig:dGdY}, also for the central model~0 (solid and dotted 
curve), which clearly exhibits the two glitches induced by \HeI\ and 
\HeII\ ionization, and which we formally represent by the sum of two Gaussian 
functions about the acoustic depths $\tau=\tau_{\rm I}$ (\HeI) and 
$\tau=\tau_{\rm II}$ (\HeII):
\begin{eqnarray}
{\partial\gamma\over\partial Y}\Big\vert_{Y=0}
\hspace{-7pt}\simeq\hspace{-3pt}-{\gamma\over\sqrt{2\pi}Y}
&&\hspace{-20pt}
\left[{\Gamma_{\rm I}\over\Delta_{\rm I}}\,{\rm e}^{-(\tau-\tau_{\rm I})^2/2\Delta^2_{\rm I}}\!+\!
{\Gamma_{\rm II}\over\Delta_{\rm II}}\,{\rm e}^{-(\tau-\tau_{\rm II})^2/2\Delta^2_{\rm II}}
\right],\cr
&&
\label{eq:2Gau}
\end{eqnarray}
in which $\gamma$ and $Y$ are the actual values in the original model.
The dashed curve in Fig.~\ref{fig:dGdY} is a least-squares fit of the two 
Gaussian functions to $\partial\gamma/\partial Y\vert_{Y=0}$ of the central 
stellar model~0. \\
The fitting parameters for the nine reference models are illustrated in
Fig.~\ref{fig:gamma2_nine-fit}. Note that the ratios 
$\beta=\Gamma_{\rm I}\Delta_{\rm II}/\Gamma_{\rm II}\Delta_{\rm I}$, 
$\eta=\tau_{\rm I}/\tau_{\rm II}$ and 
$\mu=\Delta_{\rm I}/\Delta_{\rm II}$ hardly vary. Indeed, they vary 
at most by about 35\% amongst adiabatically stratified model
envelopes whose masses and radii vary by factors of five. Therefore,
for the purpose of including \HeI\ ionization, we regard them as 
being constant. We adopt the values $\beta=0.45$ and $\eta=0.7$. But 
as is the case of several of the other parameters, particularly 
$\Delta_{\rm II}$, we have found 
that the value of $\mu$ that gives the best fit to the frequency data 
is not the same as the one that we have attempted to measure directly
from the models. This is no doubt a result of contamination by
hydrogen ionization, and perhaps ionization of heavy elements too. 
So instead we have adopted $\mu=0.90$ for all models, 
which gives an optimal fit. 


\begin{figure}
\centering
\includegraphics[width=0.93\linewidth]{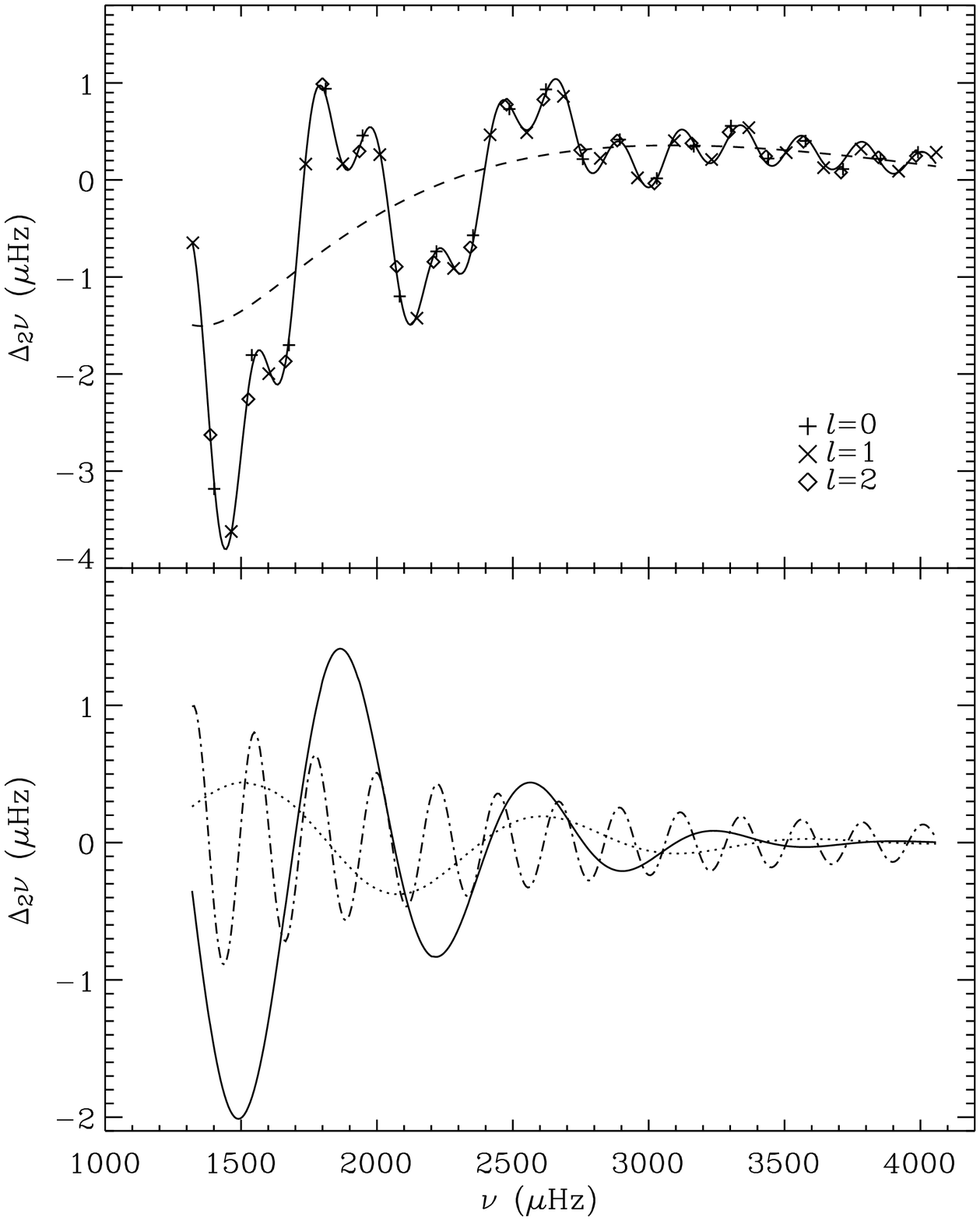}
\caption{ Top: The symbols are second differences $\Delta_2\nu$, defined by 
equation~(\ref{eq:secdiff}), of low-degree ($l$=0,1,2) eigenfrequencies 
obtained from adiabatic pulsation calculations of the central model~0, and 
have the same relation to $l$ as in Fig.~\ref{fig:GOLF}. The solid curve is 
the diagnostic $D_2$ determined by equations~(\ref{eq:sd1a}), (\ref{eq:29}) 
\& (\ref{eq:ff3a}), whose eleven parameters $\alpha_k$ have been adjusted to
fit the data by least squares. The measure $\chi^2$ (mean squared differences) 
of the overall misfit is (53\,nHz)$^2$.
The dashed curve represents the smooth contribution (last 
term in equation\,(\ref{eq:sd1a})). 
Bottom: Individual contributions of the oscillatory seismic 
diagnostic. The solid curve displays the
\HeII\ contribution, the dotted curve is the \HeI\ contribution and the
dot-dashed curve is the contribution from the base of the convection zone.
}
\label{fig:secdiff_ff3_m0}
\end{figure}


\begin{figure}
\centering
\includegraphics[width=0.93\linewidth]{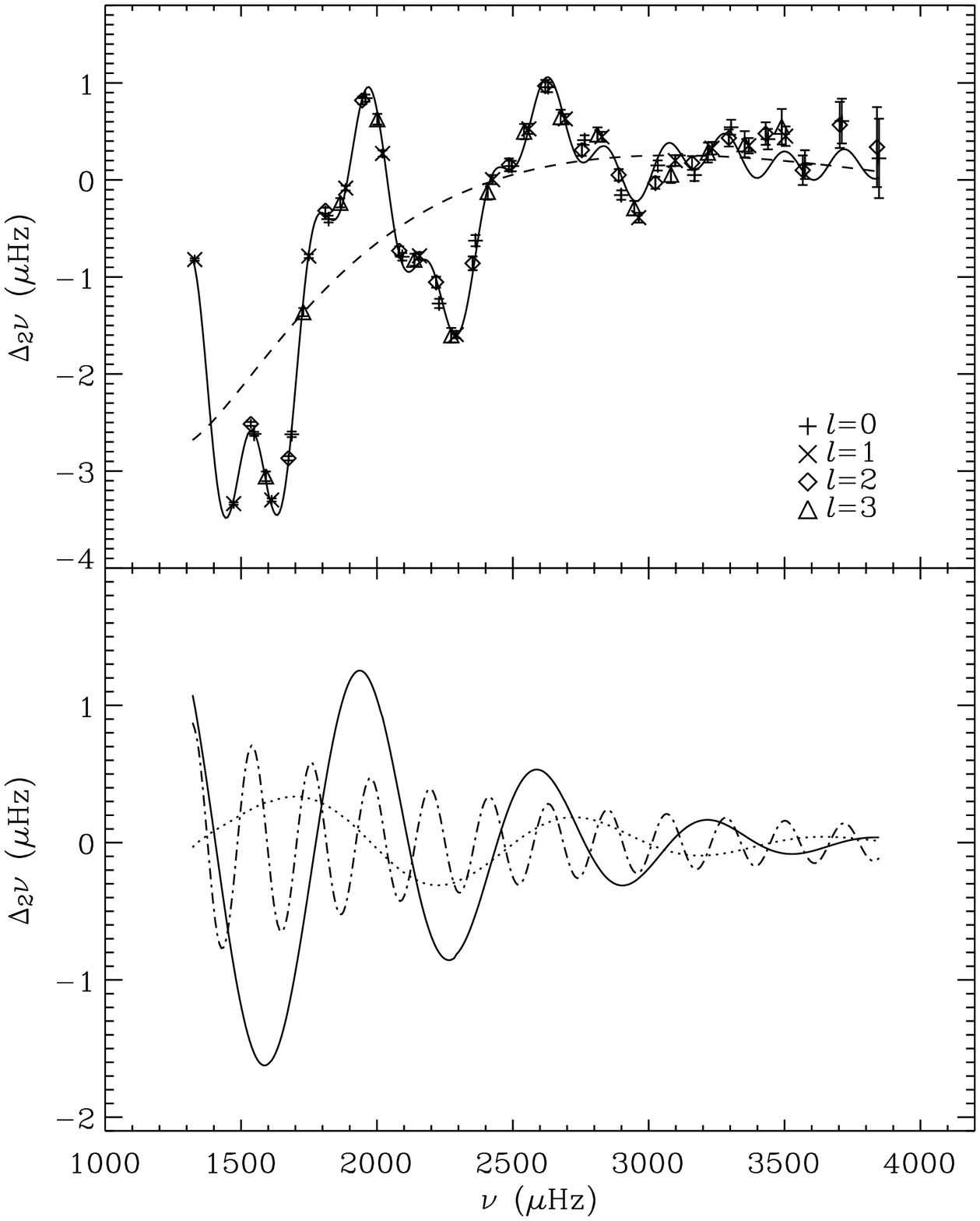}
\caption{Top: The symbols (with error bars computed under the assumption that
the raw frequency errors are independent) 
represent second differences $\Delta_2\nu$,
defined by equation~(\ref{eq:secdiff}), of low-degree solar
frequencies with $l$=0,1,2 and 3, obtained from BiSON \citep{bce06}. The
effective overall error in the data is $\langle\sigma\rangle=$5.3\,nHz.
The solid curve is the diagnostic\,$D_2(\nu;\alpha_k)$,
which has been fitted to the data in a manner intended to provide an 
optimal estimate of the eleven parameters $\alpha_k$.
The values of some of these fitting parameters are:
$-\delta\gamma/\gamma\vert_{\tau_{\rm II}}\simeq0.047$, 
$\tau_{\rm II}\simeq819\,$s, $\Delta_{\rm II}\simeq70\,$s, 
and the measure $E$ of the overall misfit is 33\,nHz. The direct measure 
$\overline\chi$ is 2.1; the minimum-$\chi^2$ fit of the function $D_2$ to 
the data yields $\overline\chi_{\rm min}=1.6$.
The dashed curve represents the smooth contribution (last term in 
equation\,(\ref{eq:sd1a})). 
Bottom: Individual contributions of the seismic diagnostic. The solid curve 
displays the \HeII\ contribution, the dotted curve is the \HeI\ contribution 
and the dot-dashed curve is the contribution from the base of the convection 
zone.  
}
\label{fig:fit_BiSON}
\end{figure}


The asymptotic evaluation of the contribution to $\delta\omega$ from
\HeI\ ionization proceeds analogously to the treatment of \HeII,
except that now one must recognize that for low-frequency modes the upper 
turning point $\tau=\tau_{\rm t}$ can be within the \HeI\ ionization zone
(in the case of the Sun and similar stars, this is not the case of
\HeII\ ionization), and the evanescence of the eigenfunction above the
turning point must be taken into account. This can be achieved via the
usual Airy-function representation. But for the purpose of evaluating the
integral $\delta_\gamma{\cal K}$ it is adequate simply to use the
appropriate high-$|\psi |$ sinusoidal or exponential asymptotic 
representations either side of the turning point to estimate the 
`oscillatory' component of the integrand, which amounts to setting
\begin{equation}
[({\rm div}\xi)^2]_{\rm osc} \simeq - \frac{\omega^3}{2\gamma pcr^2\kappa}\cos
2\psi~~,~~~~ {\rm for}~ \tau > \tau_{\rm t}\,,
\label{eq:high-psi-approx1}
\end{equation}
with $\psi$ given by equation~(\ref{eq:phi1}), without the 
subscripts~{\small II}, and $\kappa$ by equation~(\ref{eq:phi2}). Despite 
the vanishing of $\kappa$ at $\tau=\tau_{\rm t}=(m+1)/\omega$, 
expression~(\ref{eq:high-psi-approx1}) is finite at $\tau=\tau_{\rm t}$ 
because $2\psi(\tau_{\rm t})= \pi/2$. One can treat the evanescent region 
similarly, avoiding the singularity and making the
representation continuous at $\tau=\tau_{\rm t}$ by writing
\begin{equation}
[({\rm div}\xi)]^2_{\rm osc}\simeq-\frac{\omega^3}{2\gamma pcr^2|\kappa|}
       (1-{\rm e}^{2\psi -\pi/2})\,,~~{\rm for}~\tau<\tau_{\rm t}\,,
\label{eq:high-psi-approx2}
\end{equation}
where now
\begin{equation}
\psi(\tau)\simeq|\kappa|\tau\omega
    -(m+1)\ln\left(\frac{m+1}{\tau\omega}
    +|\kappa|\right) +\frac{\pi}{4}\;.
\label{eq:high-psi-approx3}
\end{equation}
Although this expression has the wrong magnitude where $-\psi$ is
large, that region makes very little contribution to the integral for
$\delta_\gamma{\cal K}$. We have confirmed numerically that the 
formula\,(\ref{eq:high-psi-approx2}) provides a tolerable approximation. 
At any point the integrand for $\delta_\gamma{\cal K}$ is the product 
of an exponential and a slowly varying function $\tilde{F}(\tau)$, say. 
Both $\tilde{F}(\tau)$
and $\psi(\tau)$ can be Taylor expanded about 
$\tau=\tilde\tau_{\rm I}=\tau_{\rm I}+\omega^{-1}\epsilon_{\rm I}$ 
($\epsilon_{\rm I}=\epsilon_{\rm II}$) up to the
quadratic term, rendering the approximation asymptotically integrable
in closed form. The correction to the result of assuming
$\tilde{F}(\tau)=\tilde{F}(\tilde\tau_{\rm I})$ and $\psi=\psi_{\rm I}$ is 
small, so we actually adopt just the leading term.

\begin{figure}
\centering
\includegraphics[width=0.93\linewidth]{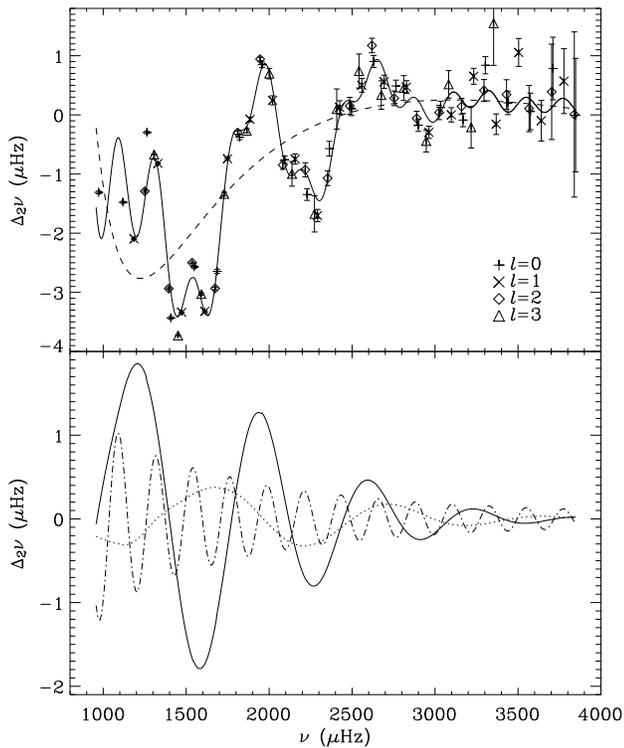}
\caption{Similar to Fig.~\ref{fig:fit_BiSON}, but with the GOLF data that 
are plotted in Fig.\ref{fig:GOLF}.
Top: The solid curve is the optimally fitted diagnostic $D_2(\nu;\alpha_k)$,
with $-\delta\gamma/\gamma\vert_{\tau_{\rm II}}\simeq0.053$, 
$\tau_{\rm II}\simeq810\,$s, and $\Delta_{\rm II}\simeq74\,$s.
The measure $E$ of the misfit is 49\,nHz, and $\overline\chi=$10.5;
the minimum-$\chi^2$ fit of $D_2$ to the data yields 
$\overline\chi_{\rm min}=7.0$. The dashed curve is the smooth contribution.
Bottom: Individual contributions to the seismic diagnostic, with the same
notation as Fig.~\ref{fig:fit_BiSON}.
}
\label{fig:fit_GOLFD2}
\end{figure}


The cyclic-eigenfrequency contribution
to the entire helium glitch then becomes
\begin{eqnarray}
\delta_\gamma\nu&\simeq&
A_{\rm II}\left[\nu+{\textstyle\frac{1}{2}}(m+1)\nu_0\right]\cr
&&\hspace{-8pt}
\times\,\Bigl[
   \mu\beta\kappa_{\rm I}^{-1}
    \left({\rm e}^{-8\pi^2\mu^2\kappa^2_{\rm I}\Delta_{\rm II}^2\nu^2}\cos2\psi_{\rm I}-1\right)\cr
&&\;\;+\;\kappa_{\rm II}^{-1}\left({\rm e}^{-8\pi^2\kappa^2_{\rm II}\Delta_{\rm II}^2\nu^2}\cos2\psi_{\rm II}-1\right)
\Bigr]\,,
\label{eq:ff3a}
\end{eqnarray}
provided $\tau_{\rm t}<\tau_{\rm I}$, in which $\kappa_{\rm II}$ is
defined in Section \ref{sec:iphase} and $\kappa_{\rm I}$ is defined
analogously; 
the phase $\psi_{\rm I}$ is given by 
$\psi_{\rm I}=\psi(\tilde\tau_{\rm I})$, where $\psi(\tilde\tau_{\rm I})$ is 
given by equation (\ref{eq:phi1}), with $\tilde\tau_{\rm II}$ replaced by
$\eta\tilde\tau_{\rm II}$, and with $\kappa_{\rm II}$ replaced by
$\kappa_{\rm I}=\kappa(\tilde\tau_{\rm I})$. If $\tau_{\rm t}>\tau_{\rm I}$, 
we replace $\kappa_{\rm I}$ with $|\kappa_{\rm I}|$ and 
$\cos 2\psi_{\rm I}$ with $1-\exp(2\psi_{\rm I}-\pi/2)$, where now
$\psi_{\rm I}=\psi(\tilde\tau_{\rm I})$ is given by 
equation~(\ref{eq:high-psi-approx3}).

Some of the parameters obtained by fitting to the
numerical second differences the generalization $D_2(\nu;\alpha_k)$ 
of the asymptotic formula\,(\ref{eq:22}) obtained by using 
expression\,(\ref{eq:ff3a})
for $\delta_\gamma\nu$, and still using equation~(\ref{eq:29}) for
$\delta_{\rm c}\nu$, are plotted in Fig.~\ref{fig:ff3}.
A comparison of the fitting parameters and of $\chi^2$ between
Figs~\ref{fig:sd1ab}, \ref{fig:sd1a_delphi} and \ref{fig:ff3} shows
that the improved seismic diagnostic $D_2$ results in an even 
better fit to the adiabatically computed eigenfrequencies than does 
the seismic diagnostic $D_0$, which includes a
description of only the \HeII\ ionization, and is also better than 
the seismic diagnostic $D_1$ with only the improved treatment of 
the asymptotic phase function.

A fit of the seismic diagnostic~$D_2$ to $\Delta_2\nu$ of the 
adiabatically computed eigenfrequencies of the central stellar model\,0 
is shown in the top panel of Fig.~\ref{fig:secdiff_ff3_m0}, and a fit to
$\Delta_2\nu$ of solar-frequency observations by BiSON \citep{bce06} in 
the top panel of Fig.~\ref{fig:fit_BiSON}. It demonstrates 
how well the oscillatory contributions to the second frequency differences 
of low-degree acoustic modes are approximated by our seismic diagnostic,
at least for frequencies above 1300$\mu$Hz.
The lower panels in both figures show the individual contributions 
from \HeI\, \HeII\ and from the discontinuity in 
${\rm d}^2\ln\rho/{\rm d}r^2$ at the base of the convection zone.

It seems likely that because the ionization diagnostic, and also the signature
of the base of the convection zone, are greatest at the
lowest frequencies, it is important for our purpose that the frequencies of 
the gravest modes of the star be measured; indeed, those modes are the
least influenced by the convective fluctuations near the stellar surface,
and therefore their frequencies are in principle the most precisely defined,
although they are apt to have low amplitudes and consequently be difficult
to detect. We note that the GOLF data do extend to frequencies lower than 
the others, and with smaller standard errors. These modes dominate the 
fitting, as is evident from a comparison of the effective overall error 
$\langle\sigma\rangle$ of the modes in the entire GOLF data set and that
of the subset within the MDI frequency range (see Section~\ref{sec:seismodel});
the higher frequencies are relatively poorly determined. 
Consequently in the error-weighted fits illustrated in 
Figs~\ref{fig:GOLF} and \ref{fig:fit_GOLFD2},
the gross properties of both the ionization signature and the signature of the 
base of the convection zone are determined predominantly by the 
lowest-frequency modes. However, we must emphasize that the
asymptotic representation used to interpret the seismic diagnostic becomes
less reliable at lower frequencies, and that therefore any model calibration 
might be more severely contaminated by structural properties other than
that caused directly by He-induced ionization.

It would be unwise 
to attempt to infer $Y$ directly from the values of the fitting parameters
obtained from this investigation because the reference 
models were not computed with the most up-to-date microphysics; however, they 
are adequate for providing the partial derivatives needed for the
calibration of a more realistic model. 

One might wonder what the effect would be of including the degree dependence
of the mode frequencies on the oscillatory component of $\delta\nu$. It can 
be taken into account by
using the full leading-order asymptotic expression for the vertical component
$K$ of the wavenumber, which is given by equation\,(\ref{eq:B2}), and making
the corresponding corrections to the eigenfunctions. For the purposes of
estimating the numerators in equations\,(\ref{eq:delomegaosc}) and 
(\ref{eq:varomc}) it is adequate to ignore the buoyancy frequency and expand
$K$ just to leading order in $l$. More care must be taken in
estimating the measures ${\cal I}$ and ${\cal I}_\Psi$ of the inertia, 
because the integrals extend down to the lower turning points of the modes
where the degree-dependent term in $K$ is as important as the other terms.
The overall effect on $\delta_{\rm osc}\nu$ is quite small, although, as we
discuss in the next section, the $l$-dependence is not entirely negligible. 
Indeed, by including the degree-dependence in the fitting procedure it is 
possible to reduce $\chi^2$ further, but only by about 10 per cent. The 
improvement, as can be seen in a plot analogous to 
Fig.~\ref{fig:secdiff_ff3_m0}, is noticeable only at low frequencies, for
which the lower turning points are higher in the star and have a larger impact
on the modes. However, the resulting modifications 
to the parameters $\delta\gamma/\gamma\vert_{\tau_{\rm II}},\ \tau_{\rm II},
\ \Delta_{\rm II}$ and $\tau_{\rm c}$ are tiny. Therefore we pursue this
matter no further.

The results of this investigation indicate that the asymptotic expression 
for $\Delta_2\nu$ can be calibrated against observation to obtain tolerable
estimates of aspects of the effect on $\gamma$ of helium ionization. They 
therefore augur the success of a future asteroseismic calibration.

   \section{Discussion}
   \label{sec:discussion}


\begin{figure*}
\centering
\includegraphics[angle=90,width=0.76\linewidth]{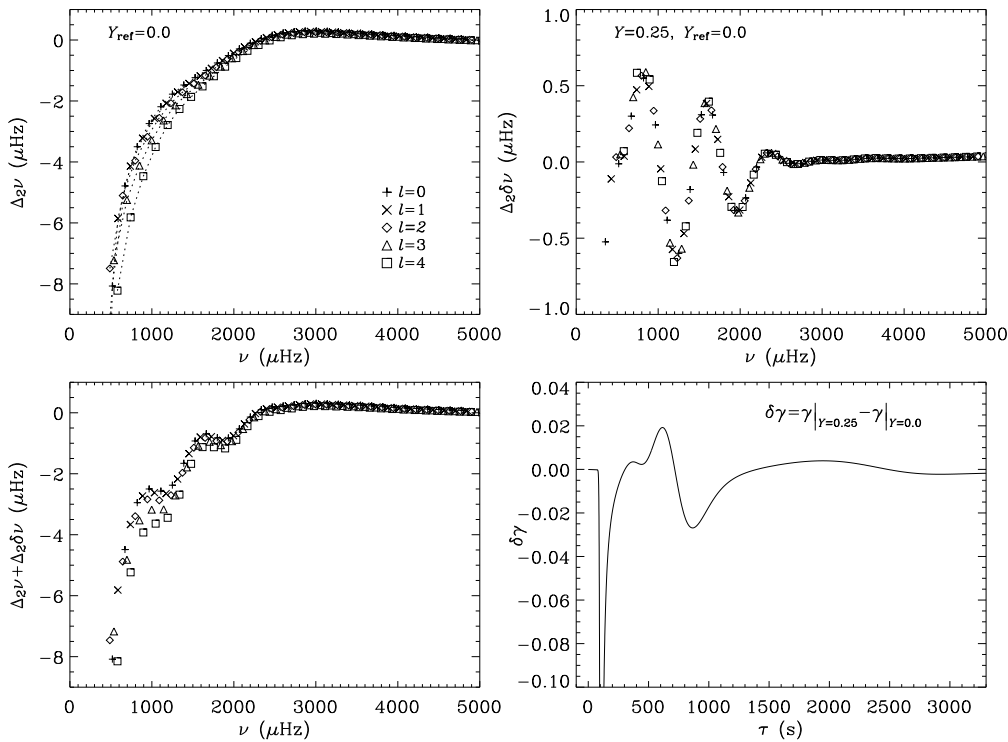}
\caption{
Calculated second frequency differences for adiabatically stratified models. 
The upper left panel shows the results for a reference model with 
helium-free gas ($Y_{\rm ref}=0.0$, $Z=0.02$). The upper right panel shows 
the second differences of the frequency perturbation $\delta\nu$ according to 
equations\,(\ref{eq:delomegaosc})--(\ref{eq:inertia}) 
($\delta\nu=\delta\omega/2\pi$) with 
$\delta\gamma=\gamma\vert_{Y=0.25}-\gamma\vert_{Y=0.0}$ being the difference 
in $\gamma$ between the reference model computed with $Y_{\rm ref}=0.0$ and 
a model with $Y=0.25$; $\delta\gamma$ is depicted in the lower right panel. 
The sum of $\Delta_2\,\nu$ and 
$\Delta_2\,\delta\nu$ is plotted in the lower left panel; it is an 
estimate of the second frequency differences of the $Y=0.25$ model. 
The meanings of the symbols in the first three panels are indicated in the
first panel.  All results are obtained with the numerically computed 
eigenfunctions of the (unperturbed) reference model.
}
\label{fig:admodels}
\end{figure*}


Although we have decided to ignore the degree-dependence in our fitting,
it is at least instructive to enquire how the degree $l$ influences the
ionization signature in the second differences. To this end we study a model 
in which we omit the effects of the base of the convection zone, accounting for
contributions to the second frequency differences from only the ionization 
zones. Such conditions are represented by an adiabatically stratified stellar
model satisfying the following equations:
\begin{eqnarray}
\frac{{\rm d}p}{{\rm d}r}&=&-\frac{Gm\rho}{r^2}\,,\\
\frac{{\rm d}m}{{\rm d}r}&=&4\pi r^2\rho\,,\\
\frac{{\rm d}\rho}{{\rm d}r}&=&-\frac{Gm\rho^2}{\gamma pr^2}\,,
\label{eq:admodstruct}
\end{eqnarray}
where $m$ is the mass enclosed in the sphere of radius $r$ (not to be
confused with the polytropic index which it represents elsewhere in
this paper). In constructing the model we adopted
$\gamma=\gamma(\rho,p,Y,Z)$ obtained from the EFF equation of state
\citep{eff73}.  The usual regularity condition $m/r^2\rightarrow0$ as
$r\rightarrow0$ and the boundary condition $\rho=\rho_{\rm c}$ were
adopted at the centre, and the conditions $m=M, p=p_{\rm s}$ 
were imposed at the surface $r=R$,  
where $p_{\rm s}$ is obtained from model S of \citet{jcd96}. 
There results an eigenvalue problem for the eigenvalue $\rho_{\rm c}$.

First, we concentrate on the smooth contributions to the second frequency 
differences. We represented them by a third-degree polynomial in 
Sections\,\ref{sec:seismodel} and \ref{sec:fimprov}, as in the last term 
in equation\,(\ref{eq:sd1a}). The symbols in the upper left panel of 
Fig.~\ref{fig:admodels} show second differences $\Delta_2\nu$ of adiabatically 
computed eigenfrequencies for a helium-free adiabatically stratified reference 
model ($Y_{\rm ref}=0$, $Z=0.02$). The dotted lines connect symbols of like 
degree $l$.
They differ only at the lowest frequencies, and by a relatively small amount
(values for $l=4$ lie the furthest from a single curve, as expected,
but modes of such high degree are not considered elsewhere in this paper). 
In the upper right panel are plotted the second differences 
$\Delta_2\delta\nu$ of the oscillatory 
component associated with helium ionization of a model in which the helium
abundance $Y$ is 0.25 ($Z=0.02$). They were obtained from evaluating numerically
the formulae\,(\ref{eq:delomegaosc})--(\ref{eq:inertia}) obtained from the 
variational principle,
with $\delta\gamma=\gamma\vert_{Y=0.25}-\gamma\vert_{Y=0.0}$ and with the 
numerically computed eigenfunctions $\bxi$ (the depth-dependence
of $\delta\gamma$, plotted as a function of acoustic depth $\tau$, 
is illustrated in the lower right panel of Fig.~\ref{fig:admodels}),
and are essentially independent of degree $l$.
With the help of equation\,(\ref{eq:delomega}) the
second frequency differences of the $Y=0.25$ model can be obtained by taking
the sum of $\Delta_2\nu$ of the $Y=0$ model and $\Delta_2\delta\nu$. The
result is depicted in the lower left panel of Fig.~\ref{fig:admodels}.
Once again the values associated with $l=0,1,2$ and 3 fall almost on a
single curve. Therefore ignoring $l$ is quite a good procedure, as is
indeed evident from the fit to the second differences of the frequencies of
the solar model presented in Fig.~\ref{fig:secdiff_ff3_m0}. That
fit is not perfect, however, and it may be that it could be improved by
extracting the $l$-dependence from the smooth contribution using the
asymptotic relation\,(\ref{eq:new19}). The $l$-dependence of the
oscillatory component is weaker; it arises principally from the 
$l$-dependence of the inertia ${\cal I}$.

On the whole, the amplitudes $\delta\gamma/\gamma\vert_{\tau_{\rm II}}$ of
the acoustic glitch inferred by fitting the asymptotically derived formulae
to the frequency differences $\Delta_2\nu$ of the test models described in
Section~\ref{sec:testmodels} agree reasonably well with the
actual values in the model, especially when our preferred expression for the
phase function is used. The amplitude increases with $Y$, as it should, even
when the He ionization zones are represented by only a single Gaussian
function (at least for the model sequence in which age varies). We notice
in passing, however, that \citet{mt05} report that if the triangular
approximation is made to the depression in $\gamma$, then the inferred 
amplitude varies in the wrong sense. The other parameters in our fitting
are not reproduced so well as the amplitude. The estimated widths 
$\Delta_{\rm II}$ are rather small, which is not what one would expect at
least from Fig.~\ref{fig:dGdY}, because the actual perturbation to
$\gamma$ appears to be somewhat broader than the Gaussian approximation
used to represent it, although not by as much as the differences suggested
by comparing Figs~\ref{fig:gamma2_nine-fit} and \ref{fig:ff3}; 
it is comforting, however, that when the two helium
ionizations are treated separately, as they should be, the magnitude of the
discrepancy in $\Delta_{\rm II}$ is the least.
Also in error are the acoustic depths
of the \HeII\ ionization zones and the bases of the convection zones,
although they are better when the improved phase function is employed, as
one would expect. We suspect that the remaining errors with the improved phase
function have arisen because the actual models, and real stars too, are not
polytropes: also there is some ambiguity in obtaining an appropriate
location of the effective acoustic surface of the Sun, and hence the origin
of the coordinate $\tau$, because that depends on precisely how the layers
in the convection zone near the upper turning points influence the dynamics
of the modes, and consequently how the region adopted for the extrapolation
of $c^2$ or $\omega^2_{\rm a}$ should be chosen; moreover, the deviation of
the stratification from the polytropic behaviour influences the relation
between $\psi$ and our polytropic approximation throughout the convection zone,
resulting in the errors in $\psi_{\rm II}$ and $\psi_{\rm c}$ being different.
Finally, we note that the value of $A_{\rm c}$ inferred from the seismic 
calibration is $20\%$ too low, which is puzzling because we would expect the
asymptotic expressions used to relate the discontinuity in the second density
derivative to the amplitude of the oscillatory signature to work
rather well. Perhaps a significant source of the discrepancy is our
neglect of the influence of the discontinuity in $\omega_{\rm a}$ on the
structure of the oscillation eigenfunctions (see Appendix B).
In any case, we stress that the precision of the asteroseismic calibrations
that are planned does not rely directly on the precision of the asymptotic
expressions. Those expressions have been used solely to
design seismic signatures of certain properties of the stratification of a
star to be as free from contamination by other properties as can
reasonably be expected; their role is to enable one to interpret the results
of the calibrations, and to help in the assessment of the accuracy of the
inferences.

It is noteworthy that the slopes of the solid and dashed lines representing
the behaviour of the two different sequences of solar models in the top
three panels of Figs~\ref{fig:gamma_nine-fit}, \ref{fig:sd1ab}, 
\ref{fig:sd1a_delphi}, \ref{fig:gamma2_nine-fit} and \ref{fig:ff3} are 
not parallel. The variations of $\tau_{\rm II}$ and $\Delta_{\rm II}$ are
both small, and can be influenced by details of the upper superadiabatic
boundary layer in the convection zone, for example, which is not of principal
concern here. But one might naively expect, and indeed hope, the values of
$\delta\gamma/\gamma\vert_{\tau_{\rm II}}$ to depend almost solely on the
value of $Y$, for then the interpretation would have been simple. But that is
evidently not the case. A substantial direct influence of heavy elements on
the depression of $\gamma$ by \HeII\ ionization may be partially responsible.
But it is likely that the predominant influence comes about through the 
dependence on the specific entropy
$s$ of the adiabat deep in the convection zone, which is determined by the
($Z$-dependent) calibration of the models to the solar radius and luminosity.
We reiterate that nevertheless this result does not compromise the precision
of the asteroseismic calibration, just its interpretation. It is no surprise
that the slopes of $A_{\rm c}(Y)$ and $\tau_{\rm c}(Y)$ are different for the
two sequences of solar models, because they depend on the interfacing of the
convection zone with the radiative interior, and the latter is quite sensitive 
to the opacity, and hence to the value of $Z$ (and, therefore in a calibrated 
sequence of solar models, to $Y$).

Our investigation has been carried out with a single equation of state.
Although it can hardly be doubted that our signature of the magnitude of
the depression in $\gamma$ is quite robust, it is certainly the case that 
relating that
to $Y$ is less so. The value of $\delta\gamma/\gamma\vert_{\tau_{\rm II}}$
is determined locally by the values of $Y$ and $s$, and the magnitude of the
error in how they are related depends directly on the error in the equation 
of state at $\tau_{\rm II}$, although it is also due partly to the inaccuracy
of our representation of the glitch (see Fig.~\ref{fig:gamma_taylor}). 
The error would certainly be adequately small
for the asteroseismic calibrations of the foreseeable future. But to
determine the value of $Y$ does seem to require knowledge of $s$, which 
can be determined only by an additional diagnostic. That diagnostic might 
have been dependent on
uncertainties in conditions far from the \HeII\ ionization zone, which may
be difficult to assess. But fortunately the immediate environment of the
helium ionization zone is isentropic, so one needs to determine only a 
constant, rather than a function: given $s$ (and the gravitational 
acceleration, which hardly varies through the helium ionization zones of 
solar-like
stars, and to a high degree of precision may be  considered to be constant
too), the functional form of $\gamma$ is rather well constrained, and a simply
measurable property of it, such as $\Delta_{\rm II}$, might be used to obtain
another, different, relation between $Y$ and $s$. That would enable each of 
these two quantities to be determined. The constancy of $s$ disengages the 
calibration from uncertainties elsewhere in the star. (We expect the
value of $\tau_{\rm II}$ to be less useful for these purposes, for it
depends on conditions in the vicinity of the upper turning points of the modes;
it should, however, be useful as a diagnostic of the treatment of convection,
which determines the structure of the superadiabatic boundary layers.)
However, before such a calibration is carried out, more thorough tests of the
reliability of the inferred value of $\Delta_{\rm II}$ must be performed.
Essentially the same suggestion has been made already by \citet{mt05} 
to distinguish between equations of state, although they too offered no 
cogent reason to suggest whether it would provide a reliable test. If the 
hypothesis that we are putting forward is correct, this measure of
the specific entropy would not alone necessarily provide a robust criterion 
for selecting any particular equation of state. However, it would certainly 
limit the uncertainty in the $\delta\gamma$--$Y$ relation needed for the 
asteroseismic calibration.

We should also point out that the variation of $\gamma$ in the He ionization
zone would be influenced by the presence of a magnetic field, which we have
ignored in this study. Evidence for low-amplitude solar-cycle variations
has been seen in the raw frequencies of a combination of solar models of
both low and intermediate degree \citep[e.g.][]{gmwk91}, which relate
directly to conditions in the \HeII\ ionization zone \citep{g94}, and,
more recently, in fourth frequency differences \citep{bm04,bm06}.
Such variations could be greater in other stars, and would therefore 
noticeably contaminate the low-degree diagnostic. Unless adequately accounted
for, they would degrade any seismic calibration of $Y$.

We emphasize that our aim in this work is not to invent a functional
representation of the second frequency differences, the diagnostic of
our choice, or of any other diagnostic for that matter, with the
intention of merely being able to fit the data well. Instead, it is
principally to obtain a formula that can be related directly to
properties of the stratification of the star that we believe to be
pertinent to the abundance of helium. Although the degree to which we
succeed in fitting the data provides some measure of our success, it
is not the sole criterion guiding our choice of function. Indeed,
functions that are not obviously directly related to the
stratification but which are tailored to mimic the data should be almost
bound to fit the data better. But evidently they must be less reliable
for diagnosis. It is encouraging, however, that our function actually
does fit the frequency data better than any other that has been
tried before \citep{hg06}, particularly when the data are fitted 
over a large range of frequency, although we hasten to add that its
ability to measure the acoustic glitches is not perfect.
In this regard it is interesting to note that the fit to the GOLF data of 
$D_2$, illustrated in Fig.~\ref{fig:fit_GOLFD2}, is no better
than the fit of $D_1$, illustrated in Fig.~\ref{fig:GOLF}, yet the values 
inferred for $-\delta\gamma/\gamma\vert_{\tau_{\rm II}}$, 
$\tau_{\rm II}$ and $\Delta_{\rm II}$ are likely to be superior.

Finally, we emphasize that although our analysis is based on
asymptotic theory, the actual calibrations would be carried out using
numerically computed eigenfrequencies, and their precision would be
independent of the accuracy of the asymptotics. As is often the case
in asteroseismology, the role of the asymptotics is to motivate the
design of a calibration procedure, and to facilitate its interpretation 
in physical terms; it is not used in the calibration itself.


   \section{Summary and Conclusion}
   \label{sec:summary}

We have designed asteroseismic signatures of helium ionization which
we trust will provide a reliable basis for calibrating stars against a
suitable grid of theoretical models, using second frequency
differences
$\Delta_2\nu_{n,l}=\nu_{n-1,l}-2\nu_{n,l}+\nu_{n+1,l}$. Those
signatures can be used to determine the magnitudes, widths and locations of
the depressions $\delta\gamma$ of the first
adiabatic exponent $\gamma$ due to He ionization.

The degree to which our original simple diagnostic~(\ref{eq:22}) appears to
reproduce the magnitude and width of the depression in $\gamma$ in the
\HeII\ ionization zone can be judged by comparing the inferred values plotted
in Fig.~\ref{fig:sd1ab} with those measured from the models in 
Fig.~\ref{fig:gamma_nine-fit}. The values of the amplitudes of the depression
are overestimated by some 50\%, and the widths underestimated by a similar
amount, yielding an approximately correct magnitude of the glitch integral,
$\Gamma_{\rm II}$.
However, the slopes of the dependence of those values on helium abundance $Y$,
particularly amongst models with varying heavy-element abundance $Z$, are not
reproduced. In this regard it should be appreciated that the reason may 
actually be that our procedure for
characterizing the amplitudes and widths of the depressions in the models may
not be seismically appropriate. The variation of the acoustic depth 
$\tau_{\rm II}$ of the centre of ionization is more-or-less reproduced.
However, the 
actual magnitude appears to be discrepant; that is due in part to our crude
representation of acoustic phase, and perhaps in part to an uncertainty in 
the origin of $\tau$. Roughly half of that discrepancy is 
accounted for by our improved treatment of the phase, discussed in
Section~\ref{sec:iphase}, as can be seen in Fig.~(\ref{fig:sd1a_delphi}).
That improvement also brings the amplitude of the $\gamma$ depression almost
into agreement with the model values, but the improvement in the inferred
widths $\Delta_{\rm II}$ is only slight. We suspected that the problem with
the widths arises from our explicit neglect of \HeI\ ionization in our 
diagnostic formula,
in which we represented the two distinct helium depressions, which in reality
separately 
influence the oscillation eigenfrequencies, by a single Gaussian function.
However, we note that the variation of $\Delta_{\rm II}$ is less than the 
variation of the other ionization-related quantities, and may therefore
perhaps not be a robust diagnostic. Nevertheless, we have introduced a
separate representation of \HeI\ ionization in Section~\ref{sec:helI}, the
result of which can be assessed by comparing with model values in
Fig.~\ref{fig:gamma2_nine-fit} with the seismically inferred counterparts 
plotted in Fig.~\ref{fig:ff3}. Interestingly, the slope with respect to $Y$ 
of the
measured model values of $\tau_{\rm II}$ for the $Z$-varying sequence are
changed, whereas that of the inferred values is not. Moreover, the slope of the
amplitudes of the inferred $\gamma$ depressions in the $Z$-varying sequence 
is improved only slightly. But the magnitude of the discrepancy between the 
inferred and the actual values of $\Delta_{\rm II}$ is reduced by a factor
three.

We have found that, for solar-like stars at least, the
magnitude of $\delta\gamma$ increases approximately linearly with
helium abundance $Y$, a property which has assisted us in accounting 
for the \HeI\ ionization zone
by relating it directly to the contribution from \HeII\ ionization. 
By so doing, a somewhat more faithful overall
representation of the \HeII\ ionization is achieved by the seismic
calibration. The magnitude of the depression in $\gamma$ is not a
function of $Y$ alone, but depends also on at least one more
parameter. We argue that for a solar calibration there 
is just one, namely the specific entropy; for other stars it must depend
also on the gravitational acceleration, whose determination would be
part of the overall calibration. If the situation is indeed that simple,
then a potential asteroseismic calibration procedure is 
almost in hand.


  \subsection*{ACKNOWLEDGEMENTS}

We are very grateful to J{\o}rgen Christensen-Dalsgaard for reading an
earlier version of the paper and suggesting improvements, and to
Philip Stark for enlightening correspondence. We thank Thierry Toutain 
and Sasha Kosovichev for supplying the MDI data, J. Ballot for supplying 
the GOLF data plotted in Figs~\ref{fig:GOLF} and \ref{fig:fit_GOLFD2}, 
and Bill Chaplin for supplying the BiSON 
data plotted in Fig.~\ref{fig:fit_BiSON}.  Support by
the Particle Physics and Astronomy Research Council is gratefully
acknowledged.

 

  \appendix
  \section{Signature of rapid variation of the adiabatic exponent}
  \label{sec:varprin}

The adiabatic eigenfrequencies $\omega$ of a nonrotating star in the
Cowling approximation satisfy a variational principle which can be
written in the form \citep{c63}
\begin{equation}
\omega^2=\frac{\cal K}{\cal I}
\label{eq:A1}
\end{equation}
where
\begin{equation}
{\cal K}\!\!=\!\!\frac{1}{4\pi}
\int\!\![\gamma p({\rm div}\bxi)^2
+2(\bxi\cdot\nabla p){\rm div}\bxi
+(\bxi\cdot\nabla p)(\bxi\cdot\nabla\ln \rho)]{\rm d}V
\label{eq:A2}
\end{equation}
and
\begin{equation}
{\cal I} = \frac{1}{4\pi}\int\rho\bxi\cdot\bxi\,{\rm d}V\,,
\label{eq:A3}
\end{equation}
the integrals being over the volume of the star. Omitted from equation
(\ref{eq:A1}) is a possible surface integral; that integral
contributes to $\omega$ at most a slowly varying function (with
respect to $\omega$), which in any case in practice is small, so we have
ignored it at the outset for simplicity. By adopting the Cowling
approximation we have also ignored the contribution from the
gravitational-potential perturbation, because that too is only 
slowly varying. For
low-degree modes of high order, the integrand in $\cal K$ is dominated
by the first term (whose integral we call ${\cal{K}}_1$), except
possibly in localized regions where the equilibrium density $\rho$
varies abruptly, such as at the base of the convection zone. Therefore
the contribution to $\delta\omega$ from a region of relatively rapid
variation of $\gamma$ produced by the ionization of an abundant
element may be approximated by
\begin{equation}
\delta\omega\simeq \frac{\delta{\cal K}_1 -\omega^2 \delta \cal I}{2\omega
{\cal I}} \simeq \frac{\delta{\cal{K}}_1}{2\omega{\cal{I}}}\,,
\label{eq:A4}
\end{equation}
since $|\delta{\cal{K}}_1|\gg\omega^2|\delta{\cal{I}}|$, where
\begin{equation}
\delta{\cal K}_1 = \frac{1}{4\pi}\int (\delta\gamma)p({\rm div}\bxi)^2
{\rm d}V +\frac{1}{4\pi} \int \gamma (\delta p)({\rm div}\bxi)^2 {\rm d}V
\label{eq:A5}
\end{equation}
\begin{equation}
\hspace{5mm}=:\delta_\gamma{\cal K}_1 +\delta_p{\cal K}_1
\label{eq:A6}
\end{equation}
and
\begin{equation}
\delta{\cal I} = \frac{1}{4\pi}\int \delta\rho\,\bxi\cdot\bxi {\rm d}V~,
\label{eq:A7}
\end{equation}
in which $\delta\gamma$ and $\delta p$ are to be regarded as the
differences between $\gamma$ and $p$ in the equilibrium star and those
in a similar smooth model in which ionization is absent.
In view of the variational principle, the differences (presumed to
be small) between the values of $\bxi$ in the star and in the smooth
model do not contribute to $\delta\omega$ to leading order in small
quantities. If $\delta\gamma$ is defined to be nonvanishing only in
the ionization region where it varies rapidly with $r$, then one
expects $\delta_\gamma{\cal K}_1$ to contribute the dominant
oscillatory (with respect to frequency) component to $\omega$, because
$\delta p$, which is related to $\delta\gamma$ via its radial
derivative, varies with $r$ more slowly, as does $\delta\rho$ which in
the helium ionization zone is directly related to $\delta p$ through
the adiabatic constraint. (Note that for low-amplitude perturbations
$\delta p~,\delta\rho,~\delta\gamma$, the relation between
$\delta\rho$ and $\delta p$ does not depend on $\delta\gamma$ to
leading order.) We have confirmed numerically that this is indeed the
case.  Consequently, most of the oscillatory contribution to
$\delta\omega$, in the absence of rotation, is contained in the term
$(2 \omega {\cal I})^{-1} \delta_\gamma {\cal K}_1$.

In estimating the integrals $\cal I$ and $\cal K$ it is adequate for
our purposes to use the high-order, low-degree asymptotic expressions
for the eigenfunctions of the smooth model, and then to evaluate the
integrals asymptotically. In solar-like stars the \HeII\ ionization
zone is well inside the propagating region of the acoustic modes, and
the leading term in the Liouville-Green approximation to the solution
of the adiabatic wave equation suffices, as is described in Section~2.
That is not necessarily the case for the \HeI\ ionization zone, and 
formally one should retain the Airy-function form of the JWKB approximation.
However, for the purpose of evaluating $\cal K$, we have found it
sufficient to adopt the large-phase sinusoidal or exponential
representations either side of the upper turning point $\tau_{\rm t}$
of the mode, even up to the turning point itself, but scaling the
eigenfunction div$\bxi$ in the evanescent zone with a constant chosen
to render that eigenfunction continuous at $r=r_{\rm t}$.

Rotation influences the eigenfrequencies via the Coriolis force (with
respect to a rigid frame of reference rotating locally with the fluid)
and via centrifugal effects which both influence the oscillation
dynamics directly and distort the equilibrium structure of the star
from sphericity. One can estimate the influence with an analysis
similar to that presented here using the more general variational
principle of \citet{lo67}. In particular, one would
obtain an estimate of the oscillatory contribution to modes of like
degree $l$ and azimuthal order $m$ induced by a tachocline flow near 
the base of the convection zone of an essentially axisymmetric star. 
However, if $\omega$ is regarded as the multiplet frequency, obtained 
as the equally
weighted average of the frequencies of all the modes of like $l$ and
$n$, but varying azimuthal order, such contributions are suppressed, at
least when rotation is slow \citep[e.g.][]{g93}. Then
the multiplet frequencies
provide a good diagnostic of helium ionization. However, if the
centrifugal force is not negligible, its average effect on the
equilibrium structure of the star would need to be taken into account
in the grid of reference stellar models used for comparison.

 \section{Signature of the acoustic glitch at the base of the convection zone}
 \label{sec:delomc}

The acoustic glitch at the base of the convection zone is essentially
a discontinuity in the second derivative of the temperature,
associated with which are discontinuities in the second derivatives of
sound speed and density. That renders all three terms in the integrand
for $\cal K$, given by equation (\ref{eq:A2}), susceptible to
perturbations of the same order. Then terms must be calculated
with some care, because there can be cancellation in leading order
when the terms are combined \citep[e.g.][]{mm94}.  
It is more straightforward, however, to 
work directly (in the planar Cowling approximation)
with the single second-order differential equation for the
$r$-dependent factor $\tilde{p}(r)$ in the Lagrangian pressure
perturbation eigenfunction, for then only one quantity is
discontinuous, namely the acoustic cutoff frequency $\omega_{\rm a}$
(and, of course, the second derivative of the eigenfunction). That
equation is \citep[e.g.][]{g93}
\begin{equation}
\frac{{\rm d}^2\Psi}{{\rm d}r^2}+K^2\Psi=0\,,
\label{eq:B1}
\end{equation}
where 
$\Psi\sim r\rho^{-1/2}\,\tilde{p}$.
The vertical wavenumber is given by
\begin{equation}
K^2=\frac{\omega^2-\omega^2_{\rm a}}{c^2}-\frac{l(l+1)}{r^2}\left(1-\frac{N^2}{\omega^2}\right)\,,
\label{eq:B2}
\end{equation}
in which $N$ is the buoyancy (Brunt-V\"ais\"al\"a) frequency: 
\begin{equation}
N^2=g\left(\frac{1}{H}-\frac{g}{c^2}\right)~;
\label{eq:B3}
\end{equation}
also
\begin{equation}
\omega^2_{\rm a}=\frac{c^2}{4H^2} 
\left(1-2\frac{{\rm d}H}{{\rm d}r}\right) 
=\frac{c^2}{4H^2}
-\frac{c^2}{2}
\frac{{\rm d}^2\ln\rho}{{\rm d}r^2}~,
\label{eq:B4}
\end{equation}
where $g$ is the local acceleration due to gravity and $H=-({\rm
d}\ln\rho/{\rm d}r)^{-1}$ is the density scale height.

Equation (\ref{eq:B1}) is to be solved subject to a regularity
condition at $r=0$ and an appropriate causality condition at $r=R$. It
is written in self-adjoint form, and therefore it follows immediately
that for modes with frequencies well below the acoustic cutoff
frequency characteristic of the atmosphere, the equation 
\begin{eqnarray}
&&\hspace{-15pt}
\int^R_0 \left(\frac{{\rm d}\Psi}{{\rm d}r}\right)^2{\rm d}r
 - \frac{l(l+1)}{\omega^2}\int^R_0 \frac{N^2\Psi^2}{r^2}{\rm d}r\cr
&&\hspace{-17pt}
+\int^R_0 \left[\frac{\omega^2_{\rm a}}{{\rm c}^2}+\frac{l(l+1)}{r^2} 
\right]\Psi^2 {\rm d}r
-\omega^2\int^R_0 \frac{\Psi^2}{{\rm c}^2}\,{\rm d}r=0
\label{eq:B5}
\end{eqnarray}
defines a variational principle for $\omega^2$ amongst acceptable
functions $\Psi$; those functions are such that the boundary terms
that arise from the integration by parts to obtain the first term in
equation (\ref{eq:B5}) either vanish (at $r=0$) or can be neglected
(at $r=R$) because they contribute at most a slowly varying function
of $\omega$.
At the base of the convection zone, $r=r_{\rm c}$, 
$\omega^2_{\rm a}$ suffers a discontinuity
\begin{equation}
\Delta_{\rm c}\equiv\left[\omega^2_{\rm a}\right]_{r_{\rm c}-}^{r_{\rm c}+}\,.
\label{eq:C3}
\end{equation}

Let $\omega^2_{\rm a}$ be extrapolated smoothly from the convection
zone down into the radiative interior to define $\omega^2_{\rm as}$
with the property $\omega^2_{\rm as}\rightarrow\omega^2_{\rm a}$ as
$r\rightarrow0$. Additionally, let us define
\begin{equation}
\delta\omega^2_{\rm a}=\omega^2_{\rm a}-\omega^2_{\rm as}\,,
\end{equation}
which satisfies $\delta\omega^2_{\rm a}=0$ for $r>r_{\rm c}$,
$\delta\omega^2_{\rm a}=-\Delta_{\rm c}$ at $r=r_{\rm c}$, and
$\delta\omega^2_{\rm a}\rightarrow 0$ as $r\rightarrow 0$.  Granted
that $\delta\omega^2_{\rm a}$ is small enough for linearization to be
valid, the contribution $\delta_{\rm c}\omega^2$ to $\delta\omega^2$
from the discontinuity in $\omega^2_{\rm a}$ is then given by
\begin{eqnarray}
&&\hspace{-15pt}
\delta_{\rm c}\omega^2\int_0^R \frac{\Psi^2}{{\rm c}^2}\,{\rm d}r
+\frac{l(l+1)}{\omega^2}\int_0^R\frac{\delta N^2}{r^2}\Psi^2\,{\rm d}r\cr
&&\hspace{-15pt}
-\frac{l(l+1)\delta_{\rm c}\omega^2}{\omega^4}\int_0^R\frac{N^2}{r^2}\Psi^2\,{\rm d}r
-\int_0^{r_{\rm c}}\frac{\delta\omega^2_{\rm a}}{\rm c^2}\Psi^2{\rm d}r
\simeq0\,.
\label{eq:B8}
\end{eqnarray}
Formally it was necessary to include the second and third terms on the
left-hand side of this equation because the variational principle
holds only for perturbations in the model that preserve hydrostatic
support, and changing $\omega^2_{\rm a}$ leads to a change $\delta N^2$ in
$N^2$. However, these terms are small for low-degree high-order p
modes (and zero for radial modes).  Consequently they can safely be
ignored, yielding
\begin{equation}
\delta_{\rm c}\omega^2\simeq
\frac{\int_0^{r_{\rm c}}\delta\omega^2_{\rm a}c^{-2}\Psi^2\,{\rm d}r}
     {\int_0^Rc^{-2}\Psi^2\,{\rm d}r}
\equiv\frac{\delta_{\rm c}\cal K}{{\cal I}_\Psi}\,.
\label{eq:varomc}
\end{equation}
This formula is adequate for the purposes of studying the signature of
helium ionization. We note, however, that the linearization may not be
adequate for studying the base of the convection zone; for that
endeavour one may need to account for the perturbations to the
eigenfunctions, as did Roxburgh and Vorontsov (1994). From asymptotic
theory the high-order eigenfunctions can be approximated by
\citep[e.g.][]{g93}:
\begin{equation}
\Psi\sim\Psi_0K^{-1/2}\sin\psi\,,
\end{equation}
between the turning points $r_{\rm b}$ and $r_{\rm t}$ at which ${\cal K}$
vanishes, where
\begin{equation}
\psi=\int_r^{r_{\rm t}}K\,{\rm d}r+\frac{\pi}{4}\,;
\end{equation}
then
\begin{eqnarray}
{\cal I}_\Psi&\simeq&\int_{r_{\rm b}}^{r_{\rm t}}\Psi^2_0 c^{-2}K^{-1}\sin^2\psi \,{\rm d}r\cr
&\simeq&\frac{1}{2}\Psi^2_0\int_{r_{\rm b}}^{r_{\rm t}}(cK)^{-1}\frac{{\rm d}r}{c}\cr
&\simeq&\frac{1}{2\omega}\Psi^2_0\int_{\tau_{\rm t}}^{\tau_{\rm b}}\left(1-\frac{\omega^2_{\rm a}}{\omega^2}\right)^{-1/2}\,{\rm d}\tau\cr
&\simeq&\frac{1}{2\omega}T\Psi^2_0\,,
\label{eq:newB10oldC11}
\end{eqnarray}
in which $\tau_{\rm t}=\tau(r_{\rm t})$ and $\tau_{\rm b}=\tau(r_{\rm b})$
are the acoustic depths of the upper and lower turning points. In our 
introductory analysis in Section~2 we ignored $\omega_{\rm a}$; in that 
case the last of equations~{\ref{eq:newB10oldC11}}
follows immediately from the equation above. When $\omega_{\rm a}$ is
included in the polytropic approximation, as in Section~4.1, the
first-order correction to $\mathcal{I}_\Psi$ vanishes, and therefore
equation~(\ref{eq:newB10oldC11}) still holds. We have also ignored the 
$l$-dependence of
$K$, in the penultimate approximation in equations
(\ref{eq:newB10oldC11}), consistently with our reduction of equation
(\ref{eq:B8}) to equation (\ref{eq:varomc}). At this level of
approximation, ${\cal{I}}_\Psi$ becomes the same as the leading-order
approximation\,(\ref{eq:inertia1}) to the inertia $\cal I$ defined by 
equation (\ref{eq:inertia}) after setting $\Psi_0=\omega$.
\smallskip

\begin{figure}
\centering
\includegraphics[width=0.93\linewidth]{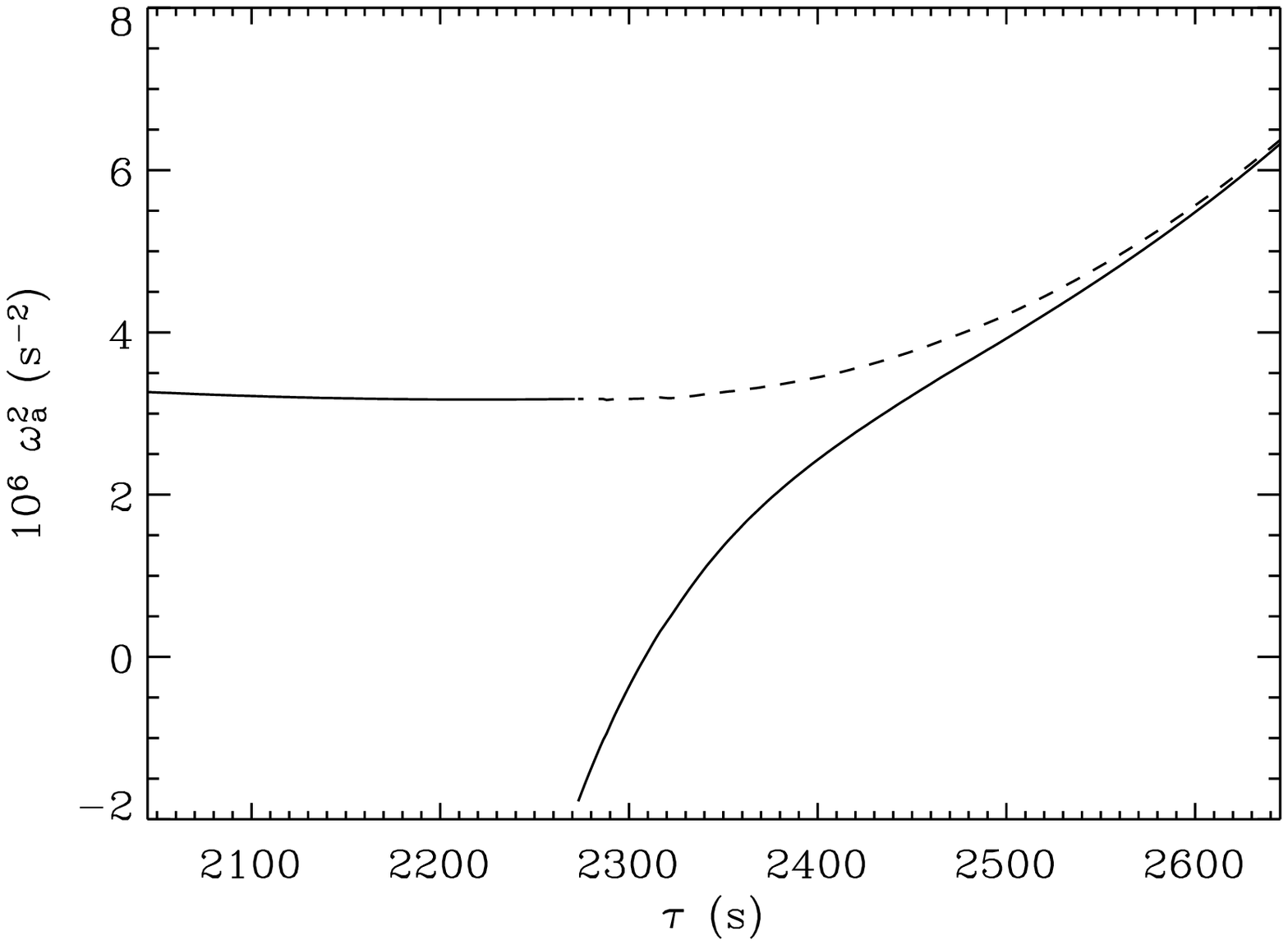}
\caption{
Acoustic cutoff frequency $\omega_{\rm a}$ in the vicinity of the base 
of the convection zone. The solid curves show $\omega^2_{\rm a}$ of
the central model\,0 and the dashed curve is the sum of the solid
curve and of equation (\ref{eq:newB13oldC14}) evaluated with
$\tau_0=80\,$s.
}
\label{fig:omega_a@bcz}
\end{figure}

\noindent The integral $\delta_{\rm c}{\cal K}$ in 
equation (\ref{eq:varomc}) can be approximated by
\begin{eqnarray}
\delta_{\rm c}{\cal K}&\simeq&\Psi^2_0\int_{r_b}^{r_{\rm c}}
     \delta\omega^2_{\rm a}c^{-2}K^{-1}\sin^2\psi\,{\rm d}r\cr
&\simeq&\frac{1}{2}\Psi^2_0\int_{r_{\rm b}}^{r_{\rm c}}
     \delta\omega^2_{\rm a}(cK)^{-2}\left[1-\cos 2\psi\right]K\,{\rm d}r\,.
\end{eqnarray}
Once again it is adequate to set $K\simeq\omega/c$, and to set
$\psi=\psi_{\rm c}+\omega(\tau-\tau_{\rm c})$ with
\begin{equation}
\psi_{\rm c}=\int_{r_{\rm c}}^{r_{\rm t}}K\,{\rm d}r+\frac{\pi}{4}\,.
\end{equation}
Furthermore, beneath the convection zone we approximate 
$\omega^2_{\rm a}$ by an exponential function:
\begin{equation}
\delta\omega^2_{\rm a}\simeq-\Delta_{\rm c}\,{\rm e}^{-(\tau-\tau_{\rm c})/\tau_0}\,,
\label{eq:newB13oldC14}
\end{equation}
where $\tau_0\ll T$ is determined by fitting expression
(\ref{eq:newB13oldC14}) to the central model of our grid; it is
compared with the model in Fig.~\ref{fig:omega_a@bcz}. 
Then, since $\delta\omega^2_{\rm a}$ varies
much more rapidly than $(cK)^{-2}$ -- in fact $cK\simeq\omega=$
constant -- one may ignore the variation of the latter over a few
$\tau_0$. It follows that the oscillatory component of 
$\delta_{\rm c}{\cal K}$ is then
\begin{equation}
\delta_{\rm c,osc}{\cal K}
\simeq\frac{1}{2}\omega\Delta_{\rm c}\Psi^2_0(cK)^{-2}\big\vert_{r_{\rm c}\!-\!}
\!\!\int_{\tau_{\rm c}}^{T}\!\!\!{\rm e}^{-(\tau-\tau_{\rm c})/
\tau_0}\cos 2\psi\,{\rm d}\tau
\end{equation}
\vspace{-10pt}
\begin{eqnarray}
\phantom{\delta_{\rm cos}{\cal K}}
&\!\!=\!\!&\frac{1}{4}\Psi^2_0(cK)^{-2}\big\vert_{r_{\rm c}\!-\!}
    \left(1+\frac{1}{4\tau^2_0\omega^2}\right)^{-1/2}\Delta_{\rm c}\cr
&&\times\cos\left[2\psi_{\rm c}+
    \tan^{-1}\left(2\tau_0\omega\right)\right]\,.
\end{eqnarray}

Finally, we notice that only the second term in equation (\ref{eq:B4})
is discontinuous, whence
\begin{equation}
\Delta_{\rm c}=-\frac{c^2}{2}
\left[\frac{{\rm d}^2\ln\rho}{{\rm d}r^2}\right]_{r_{\rm c}-}^{r_{\rm c}+}\,.
\end{equation}
Setting $(cK)^2\simeq\omega^2$ at the base of the convection zone, we
finally obtain for the oscillatory contribution $\delta_{\rm
c,osc}\omega$ to $\omega$ from the discontinuity in the acoustic
cutoff frequency at $r=r_{\rm c}$:
\begin{eqnarray}
\delta_{\rm c,osc}\omega&\simeq&\frac{c^2_{\rm c}}{8T\omega^2}
\left(1+\frac{1}{4\tau^2_0\omega^2}\right)^{-1/2}
\left[\frac{{\rm d}^2\ln\rho}{{\rm d}r^2}\right]_{r_{\rm c}-}^{r_{\rm c}+}\cr
&&\times\cos
\left[2\psi_{\rm c}+\tan^{-1}(2\tau_0\omega) \right]\,,
\label{eq:delomc}
\end{eqnarray}
where $c_{\rm c}=c(r_{\rm c})$. 

Within our initial approximation discussed in Section~2, in which the
acoustic cutoff frequency $\omega_{\rm a}$ was not explicitly taken
into account, we set $\psi_{\rm c}=\tau_{\rm c}\omega+\epsilon_{\rm c}$, 
where $\epsilon_{\rm c}$ is constant, as in equation
(\ref{eq:16}). When $\omega_{\rm a}$ is taken into account as in Section~4,
the phase is $\psi_{\rm c}=\psi(\tilde\tau_{\rm c})$, in which
$\psi(\tilde\tau_{\rm c})$ is essentially given by 
equation~(\ref{eq:phi1}) with $\tilde\tau_{\rm II}$ replaced 
by $\tilde\tau_{\rm c}=\tau_{\rm c}+\omega^{-1}\epsilon_{\rm c}$:
\begin{equation}
\psi_{\rm c}\simeq\kappa_{\rm c}\tilde\tau_{\rm c}\omega
-(m+1)\cos^{-1}\left(\frac{m+1}{\tilde\tau_{\rm c}\omega}\right)
+\frac{\pi}{4}\,,
\label{eq:newB20}
\end{equation}
where $\kappa_{\rm c}=[1-(m+1)^2/\tilde\tau^2_{\rm c}\omega^2]^{1/2}$,
which is tantamount to adding a frequency-dependent phase term 
to the constant $\epsilon_{\rm c}$. 
For stars with acoustically deep convection zones, the correction to the
initial approximation is small, and the frequency dependence is
therefore weak. In the case of the sun, for example, 
$\tilde\tau_{\rm c}\omega \simeq 40$ for a 3~mHz mode, whence 
${\rm d}\psi_{\rm c}/{\rm d}\omega=
\kappa_{\rm c}\tilde\tau_{\rm c}\simeq 0.995 \tilde\tau_{\rm c}$.

\section{Second difference of the oscillatory frequency component}
\label{sec:DOGcallsthisAppxC}
Consider an oscillatory signal of the form
\begin{equation}
\delta\nu_n=A_n\cos x_n
\label{eq:DOG_C1}
\end{equation}
with $\nu_n=\omega_n/2\pi$, where $A_n=A_n(\omega_n)$ 
and where $x_n=x_n(\omega_{n})$ varies almost linearly 
with $\omega_n$ throughout the range of
frequencies $\omega_n$ that we are considering. For clarity we have
suppressed the index $l$. To estimate the second frequency difference
$\Delta_2\delta\nu_n\equiv\delta\nu_{n+1} -2\delta\nu_n
+\delta\nu_{n-1}$ we regard $\omega_n$ as a continuous function of $n$
and expand $x_{n\pm 1}$ and $A_{n\pm 1}$ in Taylor series about $x_n$
and $A_n$ respectively, retaining only two terms for $x_{n\pm 1}$ and
three for $A_{n\pm 1}$:
\begin{equation}
x_{n\pm 1}\simeq x_n \pm \alpha~,~~ A_{n\pm 1}\simeq (1\pm
a+b)A_n~,~~ n\gg 1~,
\label{eq:DOG_C2}
\end{equation}
where  $\omega_0=2\pi\nu_0$ approximates the large (angular) frequency
separation and 
\begin{equation}
\alpha=\frac{{\rm d}\omega_n}{{\rm d}n} \frac{{\rm d}x_n}{{\rm
d}\omega_n} \simeq \omega_0\frac{{\rm d}x_n}{{\rm d}\omega_n}\,,
\label{Cnonumber}
\end{equation}
\begin{eqnarray}
a&=&\frac{1}{A_n} \frac{{\rm d}A_n}{{\rm d}n} 
 \simeq \omega_0\frac{{\rm d}\ln A_n}{{\rm d}\omega_n}\,,\cr
b&=&\frac{1}{2A_n} \frac{{\rm d}^2A_n}{{\rm d}n^2}
  = \frac{1}{2A_n} \left[\left(\frac{{\rm d}\omega_n}{{\rm d}n}\right)^2 
 \frac{{\rm d}^2A_n}{{\rm d}\omega^2_n}+\frac{{\rm d}^2\omega_n}{{\rm d}n^2} 
 \frac{{\rm d}A_n}{{\rm d}\omega_n}\right]\cr
&\simeq&\frac{\omega^2_0}{2A_n} \frac{{\rm d}^2A_n}{{\rm d}\omega^2_n}\,.
\label{eq:DOG_C3}
\end{eqnarray}
The derivatives of $\omega_n$ are obtained by differentiating equation
(\ref{eq:new19}) with respect to $n$ at fixed $l$; it is consistent
with the rest of this analysis to retain only the leading term: 
${\rm d}\omega_n/{\rm d}n \simeq \omega_0,~{\rm d}^2 
\omega_n/{\rm d}n^2\simeq 0$. 
Substituting equations (\ref{eq:DOG_C2}) into the
definition of the second difference yields
\begin{eqnarray}
\Delta_2\delta\nu_n&=& [(1+a+b)\cos (x_n+\alpha)-2\cos x_n\cr 
                   && +(1-a+b)\cos(x_n -\alpha)]A_n \cr
                   &=& FA_n\cos (x_n-\delta)\,,
\label{eq:DOG_C4}
\end{eqnarray}
where
\begin{equation}
F=2\{[1-(1+b)\cos \alpha]^2 + a^2\sin ^2 \alpha\}^{1/2}
\label{eq:DOG_F}
\end{equation}
and
\begin{equation}
\delta=\tan^{-1}\left[ \frac{a\sin \alpha}{1-(1+b)\cos\alpha}\right]\,. 
\label{eq:DOG_del}
\end{equation}
Had the frequency separation $\Delta_1\nu_n$ been small compared
with the scale of variation on $A_n$, then $\Delta_1\delta\nu_n$
and $\Delta_2\delta\nu_n$ would approximate the first and second
derivatives of $\delta\nu_n$ with respect to $n$; but it is not,
and the formulae\,(\ref{eq:DOG_C4})--(\ref{eq:DOG_del}) differ
substantially from the second derivative.

In the first approximation to the contribution\, from\, \HeII\,
ionization given by equation (\ref{eq:14}), $A_n\propto \omega_n
{\rm e}^{-2\Delta^2_{\rm II}\omega^2_n}$ and 
$x_n=2(\tau_{\rm II}\omega_n+\epsilon_{\rm II})$. Therefore, 
\begin{equation}
\alpha\simeq2\omega_0\tau_{\rm II}\,,
\label{eq:alpha_II}
\end{equation}
and
\begin{eqnarray}
     a&\simeq&-(4\Delta^2_{\rm II}\omega_n^2-1) \omega_0 \omega^{-1}_n\,,\cr
     b&\simeq&2\Delta^2_{\rm II}\omega^2_0 (4\Delta^2_{\rm II} \omega^2_n -3)\,.
\label{eq:ab_II}
\end{eqnarray}
In the case of the Sun, $\omega_0\simeq8.8\times10^{-4}$s$^{-1},\;
\Delta_{\rm II} \simeq80\,$s\, and\,
$\tau_{\rm II}\simeq800\,$s; whence $\alpha\simeq1.4$. In the middle of 
the frequency range used for our calibration of solar models,
$\omega_n\simeq 2\pi \times 2.7$ mHz $\simeq 1.7\times 10^{-2}{\rm s}^{-1}$;
whence $a\simeq -0.33$ and $b\simeq0.04$. The latter is so small partly
because of a near cancellation of the two terms in parentheses at this
frequency (cancellation is exact at $\nu\simeq2\,$mHz), but it 
never exceeds this value by more than a factor
four in the frequency range of interest, and is greatest at the
highest frequency where the amplitude $A_n$ is the smallest. The
effect of the neglected higher-order terms in the Taylor expansions of
$A_n$ and $x_n$ is even smaller. The value of the amplitude factor 
$F\simeq1.7$, and $\delta\simeq-0.4$.

When the acoustic cutoff and the second-order term in the asymptotic 
eigenfrequency relation (\ref{eq:new19}) are taken into account, as 
in Section \ref{sec:iphase} , the amplitude factor for the 
ionization glitch becomes 
$A_n\propto A_{\rm II}\kappa^{-1}_{\rm II}[\omega_n +(m+1)\omega_0/2]
\exp(-2\kappa^2_{\rm II}\Delta^2_{\rm II}\omega^2_n)$ and
$x_n=2\psi_{\rm II}$, where 
$\psi_{\rm II}=\psi(\tilde\tau_{\rm II})$, with 
$\tilde\tau_{\rm II}=\tau_{\rm II}+\omega^{-1}\epsilon_{\rm II}$ and
$\psi(\tilde\tau_{\rm II})$ given by equation (\ref{eq:phi1}). 
Then
\begin{equation}
\alpha\simeq2\omega_0\tilde\tau_{\rm II}\kappa_{\rm II}\,;
\label{eq:newC7}
\end{equation}
$\kappa_{\rm II}=\kappa(\tilde\tau_{\rm II})$ is given by 
equation\,(\ref{eq:phi2}), and
\begin{eqnarray}
a&\simeq&-[4\Delta^2_{\rm II}\omega^2_n+\kappa^{-2}_{\rm II}-1-(1+\beta_{\rm a})^{-1}]\omega_0\omega^{-1}_n\,,\cr
b&\simeq&2\Delta^2_{\rm II}\omega^2_0[4\Delta^2_{\rm II}\omega^2_n-1-2(1+\beta_{\rm a})^{-1}]\,,
\label{eq:newC8}
\end{eqnarray}
where $\beta_{\rm a}=\frac{1}{2}(m+1)\omega_0\omega^{-1}_n$.
We have simplified the expression for $b$ by ignoring a term
$(m+1)^2/4\Delta^2_{\rm II}\tilde\tau^2_{\rm II}\omega^4_n$, and other
yet smaller terms, which typically augment the already small $b$ 
by only about 1~per cent.
For the \HeI\ contribution, one must replace $\kappa_{\rm II}$ and
$\Delta_{\rm II}$ by $\kappa_{\rm I}$ and $\Delta_{\rm I}$ when
$\tau_{\rm t}<\tau_{\rm I}$. When $\tau_{\rm I}$ is in the evanescent
zone of the mode, so that $\delta\nu_n=A_n\exp(-x_n)$,
then $\Delta_2\delta\nu_n=FA_n\exp(-x_n)$, where
$F=2[(1+b)\cosh\alpha-1-a\sinh\alpha]$. The derivatives $a$ and
$b$ are still given by equations~(\ref{eq:newC7}) and (\ref{eq:newC8}), 
but now $\alpha\simeq2\omega_0\tilde\tau_{\rm I}|\kappa_{\rm I}|$.

At the base of the convection zone, 
$x_n=2\psi_{\rm c}+\tan^{-1}(2\tau_0\omega_n)\simeq 2\psi_{\rm c}$ 
and $\alpha\simeq 2\omega_0\tau_{\rm c}$ 
because $\kappa_{\rm c}=\kappa(\tilde\tau_{\rm c})$ is very close to unity. 
Here, $\tilde\tau_{\rm c}=\tau_{\rm c}+\omega^{-1}\epsilon_{\rm c}$.
Moreover, 
$A_n\propto \omega^{-2}_n(1+1/4\tau^2_0\omega^2_n)^{-1/2}$, whence
\begin{eqnarray}
a&\simeq&-2\omega_0\omega^{-1}_n(1+1/8\tau^2_0\omega^2_n)/(1+1/4\tau^2_0\omega^2_n)\,,\cr
b&\simeq&\omega^2_0\omega^{-2}_n(3+5/8\tau^2_0\omega^2_n)/(1+1/4\tau^2_0\omega^2_n)^2\,.
\label{eq:ab_c}
\end{eqnarray}
Again we have simplified $b$, this time by ignoring $1/48\tau^4_0\omega^4_n$ 
in the parentheses in the numerator, which typically contributes 
about 0.2~per cent.
In the case of the Sun, $\tau_0\simeq80\,$s and
$\tau_{\rm c}\simeq2273\,$s (see Fig.~\ref{fig:omega_a@bcz}).
Whence $a\simeq-0.10$ and $b\simeq 0.01$ in the middle of the 
frequency range, and $F\simeq 3.4$ and $\delta \simeq0.05$.

\begin{table}
\caption{
Relative signal error magnification due to order-$k$ differencing of 
frequencies, evaluated at $\nu$=1.5, 2.75 and 4.0\,mHz.
}
\label{tab:c1}
\begin{tabular}{@{}cccccccc}
\hline
   &\multicolumn{7}{c}{$e_k/F_k$} \\
\cline{2-8}
\multicolumn{8}{c}{\vspace{-5pt}}\\
\raisebox{-1.5ex}[0cm][0cm]{\hspace{7pt}$k$\hspace{7pt}} 
   &\multicolumn{7}{c}{\hspace{ 4pt}\HeII\ ionization zone
                       \hspace{26pt}Base of convection zone} \\ 
   &$\hspace{-3ex}\nu$=1.5 & 2.75 & 4.0 mHz &
   &$\hspace{-3ex}\nu$=1.5 & 2.75 & 4.0 mHz \\ 
\hline
 1 &  1.38 & 1.30 & 1.33 && 0.83 & 0.81 & 0.80 \\
 2 &  1.91 & 1.46 & 1.30 && 0.71 & 0.73 & 0.74 \\
 3 &  3.22 & 2.44 & 2.33 && 0.76 & 0.77 & 0.76 \\
 4 &  4.74 & 2.90 & 2.27 && 0.70 & 0.75 & 0.76 \\
 5 &  7.98 & 4.92 & 3.90 && 0.79 & 0.81 & 0.81 \\
 6 & 12.10 & 5.79 & 3.65 && 0.74 & 0.81 & 0.83 \\
 8 & 31.43 &11.19 & 5.57 && 0.81 & 0.90 & 0.93 \\
10 & 51.36 &21.95 & 9.27 && 0.89 & 1.02 & 1.05 \\
\hline
\end{tabular}
\end{table}

It is interesting to estimate the signal-to-noise ratio for differences
$\Delta_k\delta\nu_n$ of various orders $k$. It is straightforward to extend
the analysis above to obtain formula of the kind:
$\Delta_k\delta\nu_n=F_kA_n\cos(x_n-\delta_k)$. The frequency differences
involve a weighted sum of $k+1$ frequency increments $\delta\nu_n$, which may
be written
\begin{equation}
\Delta_k\delta\nu_n=\sum_{l=-k/2}^{k/2}d_{kl}\delta\nu_{n+l}\,,
\label{eq:hoderiv}
\end{equation}
where, for example, $d_{2l}=(1,-2,1)$ and $d_{4l}=(1,-4,6,-4,1)$.
Presuming, for simplicity, that the statistical uncertainties in the
frequency measurements are independent, all with standard deviation
$\sigma$, then the variance of the possible error in
$\Delta_k\delta\nu_n$ is ${\rm e}^2_k\sigma^2=\sum_l
d^2_{kl}\sigma^2$. Hence the magnification of the measurement error,
in units of the magnification of the magnitude of the oscillatory
signal, is $e_k/F_k$. That quotient depends on the nature and location
of the acoustic glitch and the central frequency of the contributing
modes, through $F_k$, and, through $d_{kl}$, on the order $k$ of the
difference. It is tabulated in Table\,\ref{tab:c1} for the \HeII\
ionization zone and for the base of the convection zone, at the median
frequency $\nu_n=2.75\,$mHz and the two extreme frequencies 1.5 and
4.0\,mHz.  For the \HeII\ ionization zone the signal amplitude factor
$F_k$ is augmented by a factor of about 1.4 when $k$ is increased by
unity, but that is more than offset by the augmentation, by a factor
of about 1.9, of the error magnification $e_k$.  Choosing which order
$k$ of the difference to adopt must therefore require a tradeoff
between the value of the error quotient $e_k/F_k$ and the degree of
difficulty in separating the smooth and oscillatory components. It is
interesting to note that at the base of the convection zone $e_k/F_k$
is lowest for $k=2$ (except, marginally, at the lowest frequency), 
although not greatly so, which perhaps
favours $\Delta_2\delta\nu$ as a diagnostic, in accord with the 
conclusion of Ballot et al. (2004) drawn from Monte Carlo calculations.

\section{Evaluation of the convection-zone discontinuity}
\label{sec:appendixD}
The difference equations that were used to represent the differential
equations of stellar evolution (which are based simply on second-order-accuracy 
centred differences in space and time on a fixed Lagrangian spatial mesh)
do not adequately account for the mathematical properties of the
structure in the vicinity of the base of the convection zone. The 
convective heat flux was computed from a local mixing-length formalism
with a mixing length set to a constant proportion of a pressure scale height
(which therefore does not vanish at the edge of the convection zone), which
leads to a discontinuity in the second spatial derivative of the density, and
hence, according to equation\,(\ref{eq:B4}), in the critical cutoff frequency, 
which, as is common, was implicitly presumed to be absent when adopting
the difference equations for the computations. Therefore particular care 
should be
taken when estimating conditions at the base of the convection zone for the
purpose of determining the parameters $\tau_{\rm c}$ and $A_{\rm c}$ in the
seismic signature. Perhaps the most straightforward way to have computed
the structure would have been to ensure that there was always a mesh point
at the base of the convection zone and to tailor the difference scheme
either side of that point to accommodate the correct mathematical structure
of the solutions. But the models had already been computed, and so it was
necessary instead to fit appropriate functional forms to the solutions
close, but not too close, to the base of the convection zone for the purpose
of evaluating the magnitude of the discontinuity.


\begin{figure}
\centering
\includegraphics[width=0.93\linewidth]{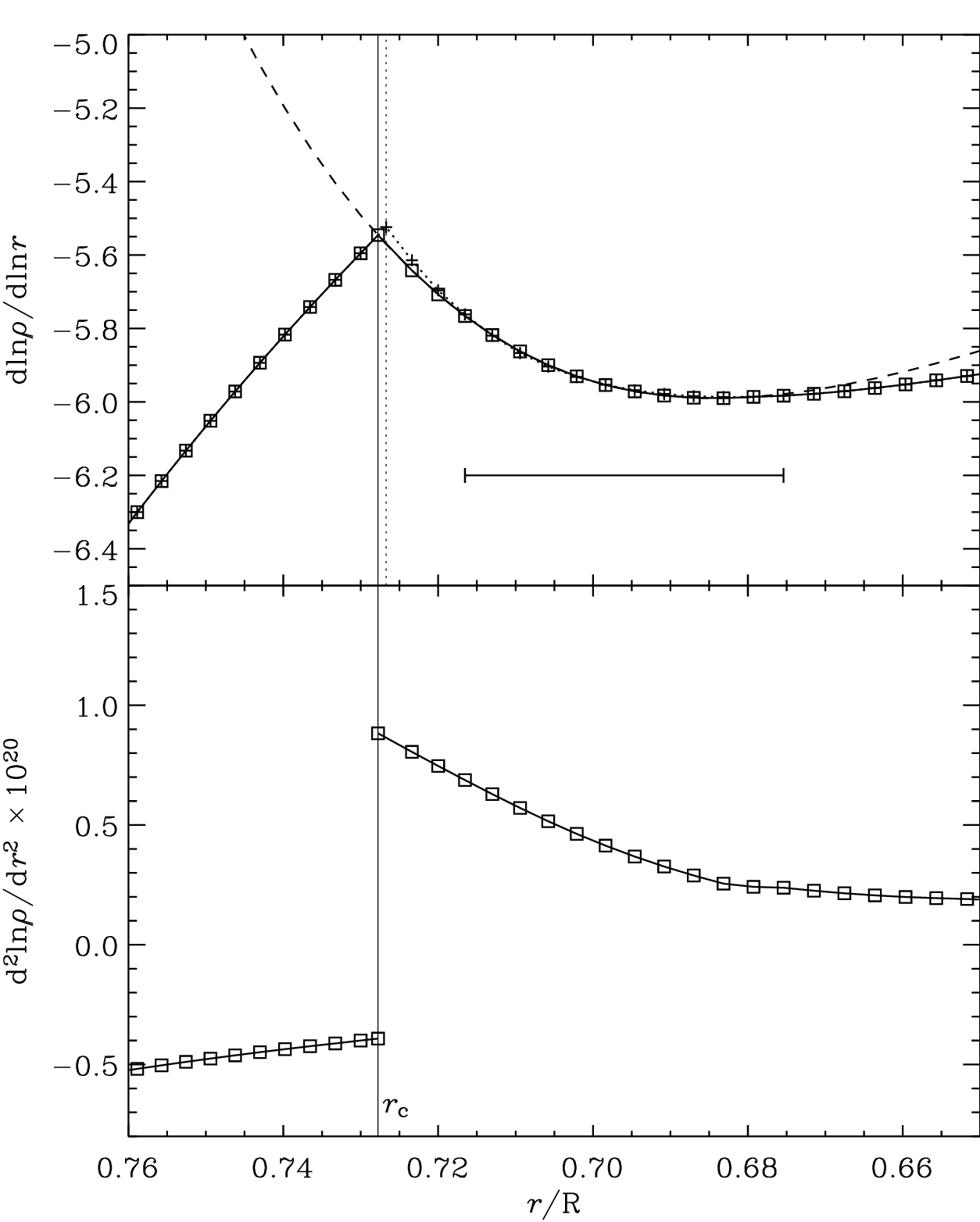}
\caption{
Density derivatives near the base of the convection zone. The top panel 
shows the first derivative of density. The plus symbols display the density 
derivatives of the original (unmodified) model and are connected by the 
dotted curve. The vertical dotted line indicates the location of the base 
of the convection zone of the unmodified model. The dashed curve is the 
result of fitting the analytical expression~\ref{eq:D1} by least 
squares to the plus symbols of the original (unmodified) model within 
the fitting domain indicated by the horizontal solid line. The square 
symbols show the density derivatives of the modified model and are connected 
by the solid curve (they lie on the dashed curve above the mid-point of the
fitting domain, and on the dotted curve below it). 
The vertical solid line indicates the new location of 
the base of the convection zone, $r_{\rm c}$. 
The bottom panel shows the result for the second density derivatives of 
the modified model obtained by numerical differentiation.
}
\label{fig:densidisc}
\end{figure}


The functional forms adopted were those derived by \citet{cgt91}. Above
the base of the convection zone, $r=r_{\rm c}$, the density variation
is smooth, and a polynomial extrapolation of ${\rm d}\ln\rho/{\rm d}r$
and ${\rm d}^2\ln\rho/{\rm d}r^2$, the latter of which was calculated by
numerical differentiation, to $r=r_{\rm c}$ is adequate. For
$r<r_{\rm c}$ we fitted to ${\rm d}\ln\rho/{\rm d}\ln r$ the function
\begin{equation}
-\frac{R}{\Lambda hr}z^{-h}
\left[1+\frac{(u-h)q}{h}z^{-u}+\frac{(2u-h)y}{h}z^{-2u}\right]^{-1}\,,
\label{eq:D1}
\end{equation}
in which $z=\rho/\rho_0$ and
\begin{equation}
\Lambda=\frac{f}{h}\frac{p_{\rm c}}{\rho_{\rm c}}\frac{R}{GM}
  \left(\frac{\rho_{\rm c}}{\rho_{\rm 0}}\right)^h
\label{eq:D2}
\end{equation}
with $p_{\rm c}=p(r_{\rm c})$, $\rho_{\rm c}=\rho(r_{\rm c})$; also
$f=(4+\lambda+v)/(3+\lambda)$, $h=f-1$, $q=\lambda hw/(1+\lambda f)$,
$y=(1+2\lambda)(f+\lambda w)hw/[2(2\lambda f+f+1)]$, $u=(1+\lambda)f$,
$w=1/(3+v)$, $\rho_0$ being an integration constant to be determined
in the fitting process, where formally 
$\lambda=(\partial\ln\varkappa/\partial\ln\rho)_{\vartheta}$ and
$v=-(\partial\ln\varkappa/\partial\ln\vartheta)_\rho$ are the 
logarithmic partial derivatives of opacity $\varkappa$ with respect
to density $\rho$ and temperature $\vartheta$, both derivatives being 
evaluated at $r=r_{\rm c}$. Expression\,(\ref{eq:D1}) was fitted 
by least squares to the solar models in the region indicated in
Fig.~\ref{fig:densidisc} by adjusting the parameters $\Lambda$, 
$\rho_0$, $\lambda$, and $v$, the outcome of which is illustrated as the 
dashed curve in the upper panel of the figure for the central model\,0.
Its intersection with the extrapolation of ${\rm d}\ln\rho/{\rm d}\ln r$ 
in the convection zone determines $r_{\rm c}$, from which it can 
then be confirmed that equation\,(\ref{eq:D2}) is satisfied. 
From this can be computed the second derivative of $\ln\rho$, 
which is illustrated in the lower panel
of Fig.~\ref{fig:densidisc}, and hence the discontinuity
${\rm d}^2\ln\rho/{\rm d}r^2\vert_{r_{\rm c}+}-
{\rm d}^2\ln\rho/{\rm d}r^2\vert_{r_{\rm c}-}$ which is needed for
evaluating $A_{\rm c}$ according to equation\,(\ref{eq:ac}). Both
$A_{\rm c}$ and $\tau_{\rm c}=\tau(r_{\rm c})$ of the nine test models
are plotted against $Y$ in the first lower two panels of 
Fig.~\ref{fig:gamma_nine-fit}.


\begin{thebibliography}{99}
\expandafter\ifx\csname natexlab\endcsname\relax\def\natexlab#1{#1}\fi
\providecommand{\enquote}[1]{``#1''}
\expandafter\ifx\csname url\endcsname\relax
  \def\url#1{\texttt{#1}}\fi
\expandafter\ifx\csname urlprefix\endcsname\relax\def\urlprefix{URL }\fi

\bibitem[\protect\citeauthoryear{Ballot, Turck-Chi\`eze\,\&\,Garc\'ia}{2004}]{btc04}
Ballot J., Turck-Chi\`eze, S., Garc\'ia R.A., 2004, A\&A, 423, 1051

\bibitem[\protect\citeauthoryear{Balmforth\,\&\,Gough}{1990}]{bg90}
Balmforth N.J., Gough D.O., 1990, ApJ, 362, 256

\bibitem[\protect\citeauthoryear{Baglin}{2003}]{bagl03}
Baglin A., 2003, Advances Space Res., 31, 345

\bibitem[\protect\citeauthoryear{Basri, Borucki \& Koch}{2005}]{bbk05}
Basri G.B., Borucki W.J., Koch D.G., 2005, New Astronomy Rev., 49, 478

\bibitem[\protect\citeauthoryear{Basu}{1997}]{b97}
Basu S., 1997, MNRAS, 288, 572

\bibitem[\protect\citeauthoryear{Basu\,\&\,Antia}{1995}]{ba95}
Basu S., Antia H.M., 1995, MNRAS, 276, 1402

\bibitem[\protect\citeauthoryear{Basu\,\&\,Mandel}{2004}]{bm04}
Basu S., Mandel A., 2004, ApJ, 617, 155

\bibitem[\protect\citeauthoryear{Basu\,\&\,Mandel}{2006}]{bm06}
Basu S., Mandel, 2006, in Lacoste H., ed., 10 years of SOHO and beyond.
ESA SP-617, Noordwijk, in press.

\bibitem[\protect\citeauthoryear{Basu, Antia~\&~Narasimha}{1994}]{ban94}
Basu S., Antia H.M., Narasimha D., 1994, MNRAS, 267, 209

\bibitem[\protect\citeauthoryear{Basu\,et\,al.}{2004}]{b04}
Basu S., Mazumdar A., Antia H. M., Demarque P., 2004, MNRAS, 350, 277

\bibitem[\protect\citeauthoryear{Basu\,et\,al.}{2006}]{bce06}
Basu S., Chaplin W.J., Elsworth Y., New A.M., Serenelli G.,
Verner G.A., 2006, ApJ, submitted

\bibitem[\protect\citeauthoryear{Baturin \& Mironova}{1990a}]{bm90a}
Baturin V.A., Mironova I.V., 1990a, Sov. Astron. Lett., 16, 108

\bibitem[\protect\citeauthoryear{Baturin \& Mironova}{1990b}]{bm90b}
Baturin V.A., Mironova I.V., 1990b, Pi\'sma Astron. Zh., 16, 253

\bibitem[\protect\citeauthoryear{Borucki\,et\,al.}{2003}]{betal03}
Borucki W.J., Koch D.G., Lissauer J.J., Basri G.B.,
Caldwell J.F., Cochran W.D., Dunham E.W., Geary J.C.,
Latham D.W., Gilliland R.L., Caldwell D.A., Jenkins J.M., 
Kondo Y., 2003, in Blades J.C., Siegmund H.W., eds,
Proceedings of the SPIE: Future EUV/UV and Visible Space 
Astrophysics Missions and Instrumentation. Vol.\,4854, p.\,129

\bibitem[\protect\citeauthoryear{Brodsky \& Vorontsov}{1987}]{bv87}
Brodsky M.A., Vorontsov S.V., 1987, Pi\'sma Astron, Zh., 13, 438

\bibitem[\protect\citeauthoryear{Chandrasekhar}{1963}]{c63}
Chandrasekhar S., 1963, ApJ, 138, 896

\bibitem[\protect\citeauthoryear{Christensen-Dalsgaard}{1981}]{cd81}
Christensen-Dalsgaard J., 1981, MNRAS, 194, 229

\bibitem[\protect\citeauthoryear{Christensen-Dalsgaard}{1986}]{cd86}
Christensen-Dalsgaard J., 1986, in Gough D.O., ed., Seismology of
the Sun and distant Stars. Nato ASI C169, Reidel, 
Dordrecht, p.\,23

\bibitem[\protect\citeauthoryear{Christensen-Dalsgaard \& Gough}{1980}]{cdg80}
Christensen-Dalsgaard J., Gough D.O., 1980, Nat, 288, 544

\bibitem[\protect\citeauthoryear{Christensen-Dalsgaard \& Gough}{1981}]{cdg81}
Christensen-Dalsgaard J., Gough D.O., 1981, A\&A, 194, 176

\bibitem[\protect\citeauthoryear{Christensen-Dalsgaard \& Gough}{1984}]{cdg84}
Christensen-Dalsgaard J., Gough D.O., 1984, in Ulrich R.K., Harvey J., 
Rhodes Jr E.J., Toomre J., eds,
Solar Seismology from Space. JPL Publ. 84-84, Pasadena, p.\,199

\bibitem[\protect\citeauthoryear{Christensen-Dalsgaard\,\&\,P\'erez Hern\'andez}{1992}]{phcd92}
Christensen-Dalsgaard J., P\'erez Hern\'andez F. ,1992, MNRAS, 257, 62

\bibitem[\protect\citeauthoryear{Christensen-Dalsgaard, Gough\,\&\,Thompson}{1991}]{cgt91}
Christensen-Dalsgaard J., Gough D.O., Thompson M.J., 1991, ApJ, 378, 413

\bibitem[\protect\citeauthoryear{Christensen-Dalsgaard\,et\,al.}{1985}]{cdghr85}
Christensen-Dalsgaard J., Duvall Jr T.L., Gough D.O., Harvey J.W., 
Rhodes Jr E.J., 1985, Nat, 315, 378

\bibitem[\protect\citeauthoryear{Christensen-Dalsgaard\,et\,al.}{1996}]{jcd96}
Christensen-Dalsgaard J., et\,al., 1996, Sci, 272, 1286

\bibitem[\protect\citeauthoryear{Cox\,\&\,Steward}{1970a}]{cs70a}
Cox A.N., Steward J.N., 1970a, ApJS, 19, 243

\bibitem[\protect\citeauthoryear{Cox\,\&\,Steward}{1970b}]{cs70b}
Cox A.N., Steward J.N., 1970b, ApJS, 19, 261

\bibitem[\protect\citeauthoryear{D\"appen, Gough\,\&\,Thompson}{1988}]{dgt88}
D\"appen W., Gough D.O.,  Thompson M.J., 1988, in Rolfe E.J., ed.,
Seismology of the Sun and Sun-like Stars.  ESA SP-286, Noordwijk, p.\,505

\bibitem[\protect\citeauthoryear{D\"appen\,et\,al.}{1991}]{dgkt91}
D\"appen W., Gough D.O., Kosovichev A.G., Thompson M.J., 1991, 
in Gough D.O., Toomre J., eds, 
Challenges to Theories of the Structure of Moderate-mass Stars.
Lecture Notes in Physics, Vol.\,388, Springer Verlag, Heidelberg, p.\,111

\bibitem[\protect\citeauthoryear{Eggleton, Faulkner \& Flannery}{1973}]{eff73} 
Eggleton P., Faulkner J., Flannery B.P., 1973, A\&A, 23, 325

\bibitem[\protect\citeauthoryear{Goldreich\,et\,al.}{1991}]{gmwk91} 
Goldreich P., Murray N., Willette G., Kumar P., 1991, ApJ, 370, 387

\bibitem[\protect\citeauthoryear{Gough}{1984a}]{g84a}
Gough D.O., 1984a, Mem. Soc. Astron. Ital., 55, 13

\bibitem[\protect\citeauthoryear{Gough}{1984b}]{g84b}
Gough D.O., 1984b, Phil. Trans. R. Soc. Lond., A, 313, 27

\bibitem[\protect\citeauthoryear{Gough}{1986a}]{g86a}
Gough D.O., 1986a, in Gough D.O., ed., Seismology of the Sun and 
distant Stars. Nato ASI C169, Reidel, Dordrecht, p.\,283

\bibitem[\protect\citeauthoryear{Gough}{1986b}]{g86b} 
Gough D.O., 1986b, in Osaki Y., ed.,
Hydrodynamic and Magnetohydrodynamic Problems in the Sun and Stars.
University Tokyo Press, Tokyo, p.\,117

\bibitem[\protect\citeauthoryear{Gough}{1987}]{g87}
Gough D.O., 1987, Nat, 326, 257

\bibitem[\protect\citeauthoryear{Gough}{1990}]{g90}
Gough D.O., 1990, in 
Osaki Y., Shibahashi H., eds, 
Progress of Seismology of the Sun and Stars. 
Lecture Notes in Physics, Vol.\,367, Springer Verlag, Heidelberg, p.\,283

\bibitem[\protect\citeauthoryear{Gough}{1993}]{g93}
Gough D.O., 1993, in 
Zahn J.-P., Zinn-Justin J., eds, 
Astrophysical fluid dynamics. 
Amsterdam, Elsevier, p.\,399

\bibitem[\protect\citeauthoryear{Gough}{1994}]{g94}
Gough D.O., 1994, in Pap J.M., Fr\"ohlich C., Hudson H.S., Solanki S.K., eds,
The Sun as a variable star. Proc. IAU Colloq., 143, Cambridge University Press,
Cambridge, p.\,252

\bibitem[\protect\citeauthoryear{Gough}{1998}]{g98}
Gough D.O., 1998, in 
Kjeldsen H., Bedding T., eds, 
Proceedings of the First MONS Workshop.
Aarhus Universitet, Aarhus, p.\,33
   
\bibitem[\protect\citeauthoryear{Gough}{2001}]{g01}
Gough D.O., 2001, in 
von Hippel T., Simpson C., Manset N., eds.,
ASP Conf. Ser.\,245, Astrophysical Ages and Timescales.
Astron. Soc. Pac., San Francisco, p.\,31

\bibitem[\protect\citeauthoryear{Gough}{2002}]{g02}
Gough D.O., 2002, in 
Favata F., Roxburgh I.W., Galadi D., eds, 
Stellar structure and habitable planet finding.
ESA SP-485, Noordwijk, p.65

\bibitem[\protect\citeauthoryear{Gough\,\&\,Novotny}{1990}]{gn90}
Gough D.O., Novotny E., 1990, Solar Physics, 128, 143

\bibitem[\protect\citeauthoryear{Gough\,\&\,Vorontsov}{1995}]{gv95}
Gough D.O., Vorontsov S.V., 1995, MNRAS, 273, 573

\bibitem[\protect\citeauthoryear{Grevesse}{1984}]{gr84}
Grevesse N., 1984, Physica Scripta, T8, 49

\bibitem[\protect\citeauthoryear{Grundahl\,et\,al.}{2006}]{getal06}
Grundahl F., Kjeldsen H., Frandsen S., Andersen M., Bedding T.,
Arentoft T., Christensen-Dalsgaard J., 2006, 
Mem. Soc. Astron. Ital., 77, 458

\bibitem[\protect\citeauthoryear{Houdek \& Gough}{2006}]{hg06}
Houdek G., Gough D.O., 2006, in Fletcher K., ed., 
Proc. SOHO 18/GONG 2006/HelAs~I, 
Beyond the Spherical Sun. ESA SP-624, Noordwijk, in press

\bibitem[\protect\citeauthoryear{Kjeldsen\,et\,al.}{2005}]{ketal05}
Kjeldsen H., Bedding T.R., Butler R.P., Christensen-Dalsgaard J.,
Kiss L.L., McCarthy C., Marcy G.W., Tinney C.G., Wright J.T., 2005,
ApJ, 635, 1281

\bibitem[\protect\citeauthoryear{Kosovichev}{1993}]{kos93}
Kosovichev A.G., 1993, MNRAS, 265, 1053

\bibitem[\protect\citeauthoryear{Kosovichev\,et\,al.}{1992}]{kcddgt92}
Kosovichev A.G., Christensen-Dalsgaard J., D\"appen W., Dziembowski W.A.,
Gough D.O., Thompson M.J., 1992, MNRAS, 259, 536

\bibitem[\protect\citeauthoryear{Lopes \& Gough}{2001}]{lg01}
Lopez I., Gough D.O., 2001, MNRAS, 322, 472

\bibitem[\protect\citeauthoryear{Lopes\,et\,al.}{1997}]{ltmg97}
Lopez I., Turck-Chi\`eze S., Michel E., Goupil M-J., 1997, ApJ, 480, 794

\bibitem[\protect\citeauthoryear{Lynden-Bell\,\&\,Ostriker}{1967}]{lo67}
Lynden-Bell D., Ostriker, J., 1967, MNRAS, 136, 293

\bibitem[\protect\citeauthoryear{Miglio\,et\,al.}{2003}]{mig03}
Miglio A., Christensen-Dalsgaard J., Di Mauro M.P., Monteiro M.J.P.F.G.,
Thompson M.J., 2003, in Thompson M.J., Cunha M.S., Monteiro M.J.P.F.G., eds,
Asteroseismology across the HR Diagram. 
Kluwer, Dordrecht, p.\,537

\bibitem[\protect\citeauthoryear{Monteiro\,\&\,Thompson}{1998}]{mt98}
Monteiro M.J.P.F.G., Thompson M., 1998, in 
Deubner F.-L., Christensen-Dalsgaard J., Kurtz D., eds. 
Proc. IAU Symp.\,185, New Eyes to see inside the Sun and Stars.
Kluwer, Dordrecht, p.\,317

\bibitem[\protect\citeauthoryear{Monteiro\,\&\,Thompson}{2005}]{mt05}
Monteiro M.J.P.F.G., Thompson M., 2005, MNRAS, 361, 1187

\bibitem[\protect\citeauthoryear{Monteiro, Christensen-Dalsgaard\,\&\,Thompson}{1994}]{mm94}
Monteiro M.J.P.F.G., Christensen-Dalsgaard J., Thompson M., 1994, 
A\&A, 283, 247

\bibitem[\protect\citeauthoryear{P\'erez Hern\'andez\,\&\,Christensen-Dalsgaard}{1994a}]{phcd94a}
P\'erez Hern\'andez F., Christensen-Dalsgaard J., 1994a, MNRAS, 267, 111

\bibitem[\protect\citeauthoryear{P\'erez Hern\'andez\,\&\,Christensen-Dalsgaard}{1994b}]{phcd94b}
P\'erez Hern\'andez F., Christensen-Dalsgaard J., 1994b, MNRAS, 269, 475

\bibitem[\protect\citeauthoryear{P\'erez Hern\'andez\,\&\,Christensen-Dalsgaard}{1998}]{phcd98}
P\'erez Hern\'andez F., Christensen-Dalsgaard J., 1998, MNRAS, 295, 344

\bibitem[\protect\citeauthoryear{Piau, Ballot\,\&\,Turck-Chi\`eze}{2005}]{pbt05}
Piau L., Ballot J., Turck-Chi\`eze S., 2005, A\&A, 430, 571

\bibitem[\protect\citeauthoryear{Roxburgh\,\&\,Vorontsov}{1994}]{rv94}
Roxburgh I.W., Vorontsov S.V., 1994, MNRAS, 268, 880

\bibitem[\protect\citeauthoryear{Shibahashi, Noels \& Gabriel}{1983}]{sng83}
Shibahashi H., Noels A., Gabriel M., 1983, A\&A, 123, 283

\bibitem[\protect\citeauthoryear{Tassoul}{1980}]{tass80}
Tassoul M., 1980, ApJS, 43, 469

\bibitem[\protect\citeauthoryear{Ulrich}{1986}]{u86}
Ulrich R.K., 1986, ApJ, 306, L37

\bibitem[\protect\citeauthoryear{Ulrich \& Rhodes Jr}{1983}]{ur83}
Ulrich R.K., Rhodes Jr E.J., 1983, ApJ, 265, 551 

\bibitem[\protect\citeauthoryear{Verner, Chaplin \& Elsworth}{2004}]{ver04}
Verner G.A., Chaplin W.J., Elsworth Y., 2004, MNRAS, 351, 311

\bibitem[\protect\citeauthoryear{Vorontsov}{1988}]{vor88}
Vorontsov S.V., 1988, in Rolfe E.J., ed., Seismology of the Sun and Sun-like Stars.
ESA SP-286, Noordwijk, p.\,475

\bibitem[\protect\citeauthoryear{Vorontsov, Baturin \& Pamyatnykh}{1991}]{vbp91}
Vorontsov S.V., Baturin V.A., Pamyatnykh A.A., 1991, Nat, 349, 49

\bibitem[\protect\citeauthoryear{Vorontsov, Baturin \& Pamyatnykh}{1992}]{vbp92}
Vorontsov S.V., Baturin V.A., Pamyatnykh A.A., 1992, MNRAS, 257, 32

\end{thebibliography}
\end{document}